\documentstyle[12pt,epsfig]{article}
\renewcommand{\thesubsection}{\arabic{subsection}}

\renewcommand{\theequation}{\arabic{subsection}.\arabic{equation}}
\renewcommand{\thefootnote}{*}
\begin{document}
\vspace*{0.5cm}
\begin{center} \Large
 The one-loop renormalization of the MSSM Higgs sector \\ and
 its application to the neutral scalar Higgs masses\footnote{
 Supported in part by the European Union under contract CHRX-CT92-0004}
\renewcommand{\thefootnote}{\arabic{footnote}}
\setcounter{footnote}{0}
\\[1.0cm]
 \large
 A. Dabelstein\footnote{
 E-Mail: ADD@DMUMPIWH.BITNET} \\
 \vspace{0.8cm}
 \normalsize \sl
 Institut f\"ur Theoretische Physik\\
 Universit\"at Karlsruhe\\
 Kaiserstr. 12\\
 D-76128 Karlsruhe, Germany \\
\end{center}
\rm
\vspace{0.3cm}
\begin{abstract}
\vspace*{0.1cm} \hspace{0.5cm}
The structure of the Higgs sector in the minimal supersymmetric
standard model is reviewed at the one-loop level.
An on-shell renormalization scheme of the MSSM Higgs sector
is presented  in detail
together with the complete list of formulae for the neutral Higgs masses
at the one-loop level.
The results of a complete one-loop calculation for the mass spectrum of
the neutral MSSM Higgs bosons and the quality of simpler
Born-like approximations are discussed for sfermion and gaugino masses
in the range of the electroweak scale.

\end{abstract}
\normalsize
\renewcommand{\thepage}{}
\newpage
\setcounter{page}{1}
%
%
\subsection{Introduction}
\vspace*{0.1cm} \hspace*{0.5cm}
\renewcommand{\thepage}{\arabic{page}}
The Higgs boson of the standard model is the last unobserved
particle since the recent results from FERMILAB indicate that
the top quark mass is rather heavy \cite{top}.
In spite of the success of the standard model, there are also good
reasons to consider ''new physics'' beyond the standard model.
One of these theoretical motivations is the appearance of
quadratically divergent contributions to the mass of the scalar
Higgs particle, so that the Higgs couplings become strong at the
TeV scale. \par
\smallskip
This problem of naturalness is solved in supersymmetric theories.
The minimal supersymmetric standard model (MSSM) is the supersymmetric
extension of the standard model, with a 2-Higgs doublet sector, where
the coefficients of the Higgs potential are restricted by supersymmetry
\cite{hunter}.
\par \smallskip
For the experimental search it is crucial to have precise
predictions for the properties of the Higgs bosons under
inclusion of radiative corrections, i.e. the
characteristic production and decay channels of the MSSM Higgs
particles at $e^+ e^-$ and $pp$ colliders.
These signatures may
allow to distinguish between a Higgs sector of different origins.
\par \smallskip
As a result of the supersymmetric Higgs potential, a light Higgs boson
exists with a tree level upper mass bound given by the $Z^0$ mass.
Radiative corrections to the Higgs mass spectrum, however, predict
an upper limit of the light Higgs mass $\cal{O}\rm $($130$ GeV)
\cite{hempf1, upper1}.
Calculations were performed at the one-loop level using renormalization
group technique \cite{barb1}, effective potential approximation \cite{ellis}
and one-loop calculations with top and stop contributions \cite{okada,yama}.
Two-loop effects to the upper limit of the lightest Higgs boson
mass are discussed in \cite{upper2}.
\par \smallskip
This article contains a complete on-shell renormalization scheme for
the MSSM Higgs sector and points out the different treatments of
the renormalization conditions of the vacuum expectation values $v_1$,
$v_2$ in the on-shell scheme \cite{pok1}. The complete MSSM expressions for
the 2-point functions of the MSSM Higgs sector are calculated and
formulae are listed in the appendix.
\par \smallskip
In chapter 4 the calculation of the full one-loop contribution to
the physical neutral scalar Higgs masses is discussed. The numerical
analysis of the one-loop contribution includes the full parameter
space of the MSSM, where the mass range of light sfermions ($> 100$ GeV)
and gauginos ($> 50$ GeV) was analyzed in detail. These one-loop
Higgs mass predictions are finally compared with simpler approximate
formulae. Deviations are within $ \approx 2 - 10$ GeV between the full
one-loop calculation and approximation formulae.
%
%
\subsection{The Higgs sector of the MSSM}
\setcounter{equation}{0}\setcounter{footnote}{0}
\subsubsection{Tree level structure}
\vspace*{0.1cm} \hspace*{0.5cm}
The Higgs sector of the MSSM consists of two scalar doublets
\begin{equation}
H_1 = \left( \begin{array}{c} H_1^1 \\ H_1^2 \end{array} \right) =
      \left( \begin{array}{c} (v_1 + \phi_1^{0} - i \chi_1^{0})/\sqrt{2} \\
 - \phi_1^-       \end{array}  \right)  \ , \
H_2 = \left( \begin{array}{c} H_2^1 \\ H_2^2 \end{array} \right) =
      \left( \begin{array}{c} \phi_2^+ \\
      (v_2 + \phi_2^0 + i \chi_2^0)/\sqrt{2} \end{array} \right) \ , \\
\end{equation}
with opposite hypercharge $Y_1 = -Y_2 = -1$ and vacuum expectation
values $v_1, v_2$. The quadratic part of the potential
\begin{eqnarray}
 V & = & m_1^2\ H_1 \bar{H_1} + m_2^2\ H_2 \bar{H_2} + m_{12}^2\ ( \epsilon
 _{ab} H_1^a H_2^b + h.c. ) + \nonumber \\
   & &     + \frac{1}{8}\ (g_1^2 + g_2^2)\ ( H_1
  \bar{H_1} - H_2 \bar{H_2} )^2 - \frac{g_2^2}{2}\ | H_1 \bar{H_2} |^2 \ ,
\label{hpot}
\end{eqnarray}
with soft breaking parameters $m_1^2$, $m_2^2$, $m_{12}^2$ and the
gauge couplings $g_1$, $g_2$ is diagonalized by the rotations
\begin{eqnarray}
 \left( \begin{array}{c} H^0 \\ h^0 \end{array} \right) & = &
 \left( \begin{array}{cc} \cos \alpha & \sin \alpha \\
 -\sin \alpha & \cos \alpha  \end{array} \right)
 \left( \begin{array}{c} \phi_1^0 \\ \phi_2^0 \end{array} \right)
\nonumber \\  \left( \begin{array}{c} G^0 \\ A^0 \end{array} \right) & = &
  \left( \begin{array}{cc} \cos \beta & \sin \beta \\
  -\sin \beta & \cos \beta \end{array} \right)
 \left( \begin{array}{c} \chi_1^0 \\ \chi_2^0 \end{array} \right)
\nonumber \\  \left( \begin{array}{c} G^+ \\ H^+ \end{array} \right) & = &
 \left( \begin{array}{cc} \cos \beta & \sin \beta \\
 -\sin \beta & \cos \beta \end{array} \right)
 \left( \begin{array}{c} \phi_1^+ \\ \phi_2^+ \end{array} \right) \ .
\label{rothig}
\end{eqnarray}
$G^0, G^\pm$ describe the unphysical Goldstone modes.
The spectrum of physical states consists of \\[-1.2cm]
\begin{table}[ht]
\begin{center}
\begin{tabular}{lll}
 2 neutral bosons with CP = 1: & $h^0, H^0$ & ("scalars") \\
 1 neutral boson with CP = -1: & $A^0$ & ("pseudoscalar") \\
 2 charged bosons: & $H^\pm$ & \ . \\
\end{tabular}
\end{center}
\end{table} \\[-1.3cm]
\smallskip \par
The masses of the gauge bosons and the electromagnetic charge are determined
by
\begin{equation}
 M_Z^2 = \frac{1}{4} ( g_1^2 + g_2^2 ) ( v_1^2 + v_2^2 ) \ , \
 M_W^2 = \frac{1}{4} g_2^2 ( v_1^2 + v_2^2 )
\label{gmass}
\end{equation}
$$
 e^2 = \frac{ g_1^2 g_2^2}{g_1^2 + g_2^2} \ .
$$
Thus, the potential (\ref{hpot}) contains two independent free
parameters, which can conveniently be chosen as
\begin{equation}
 \tan \beta = \frac{v_2}{v_1} \ , \
 M_A^2 = - m_{12}^2 \ ( \tan \beta + \cot \beta ) \ ,
\label{baspara}
\end{equation}
where $M_A$ is the mass of the $A^0$ boson.
\smallskip \par
Expressed in terms of (\ref{baspara}), the masses of the other physical
states read:
\begin{eqnarray}
m_{H^0 , h^0}^2 & = & \frac{1}{2} \ [ \ M_A^2 + M_Z^2 \pm \sqrt{ ( M_A^2 +
M_Z^2 )^2 - 4 M_Z^2 M_A^2 \cos^2 2\beta } \ ]  \nonumber \\
m_{H^+}^2 & = & M_A^2 + M_W^2 \ ,
\label{gl3}
\end{eqnarray}
and the mixing angle $\alpha$ in the $(H^0, h^0)$-system is
derived from
\begin{equation}
 \tan 2 \alpha = \tan 2 \beta \ \frac{ M_A^2 + M_Z^2 }{ M_A^2 - M_Z^2 } \ ,
 \ - \frac{\pi}{2} < \alpha \leq 0 \ .
\label{alphao}
\end{equation}
Hence, masses and couplings are determined by only a single
parameter more than in the standard model. \\
The dependence
on $M_A$ is symmetric under $\tan \beta \leftrightarrow 1/\tan \beta$,
and $m_{h^0}$ is constrained by:
\begin{equation}
m_{h^0} < M_Z \cos 2 \beta < M_Z \ .
\end{equation}
This simple scenario, however, is changed when radiative corrections
are taken into account.
%
%
\subsubsection{Renormalization of the MSSM Higgs sector}
\vspace*{0.1cm} \hspace*{0.5cm}
It is well known since quite some time that radiative corrections modify
the tree level relations (\ref{gl3}, \ref{alphao}) substantially, with
the main effect from loops involving the top quark and its scalar
partner $\tilde{t}$ \cite{hempf1,ellis}.
Various approaches have been applied:\\
\begin{itemize}
\item[(i)] The effective potential method \cite{ellis,okada} :
\\
The tree level mass matrix $\cal{M}_{\rm 0}$ of the neutral scalar system is
diagonalized by (\ref{rothig}). Loop contributions to the quadratic part of the
potential (neglecting the $q^2$-dependence of the diagrams) modify the
mass matrix
$$ \cal{M}_{\rm{0}} \rightarrow \cal{M}_{\rm{0}} + \delta \cal{M} = \cal{M}
\ . $$ Re-diagonalizing the one-loop Matrix $\cal{M}$ yields the corrected
mass eigenvalues $M_{H^0,h^0}$,
replacing (\ref{gl3}), and an effective mixing angle $\alpha_{eff}$ instead
of (\ref{alphao}).
\item[(ii)] The renormalization group method \cite{barb1}
\\
Solving the renormalization group equations for the parameters of a
general 2-doublet model and imposing the SUSY constraints at the
scale $\mu = M_{SUSY}$ yields the effective parameters of the Higgs
potential at the electroweak scale. Large log terms are resummed, but
effects from realistic mass spectra are not covered by this approximation.
\item[(iii)] Complete one-loop calculation \cite{pok1,dabx}: \\
A complete one-loop calculation to masses and couplings accommodates all
SUSY particles and mass parameters (or soft breaking parameters,
respectively) in the radiatively corrected version of
(\ref{gl3}, \ref{alphao}) and, in addition, provides the 3-point functions
required for Higgs boson production and decay processes. They are necessary
to check the quality of the approximations (i), (ii) and allow a detailed
study of the full parameter dependence of production cross sections and
decay rates. Complete one-loop calculations are available for: \\
-\hspace{0.1cm} 2-point functions (mass relations) \cite{pok1,dabx}
\\
-\hspace{0.1cm} 3-point functions $ZZh(H)$, $ZAh(H)$, $h(H)\gamma \gamma$,
... for production and \\ \hspace*{0.3cm} bosonic decay processes
\cite{pok6}. \\
-\hspace{0.1cm} 3-point functions $Aff$, $h(H)ff$ for fermionic decay
processes \cite{dabx}. \\
Other one-loop calculations with restrictions to the dominating
fermion-sfermion loops have been performed in \cite{yama}.\\
\end{itemize}
Loop calculations in the Higgs sector require an extension of the
renormalization procedure applied in the minimal version of the
standard model, e.g. the on-shell scheme \cite{bhs}.
\medskip \par
At the one-loop level, the free parameters and the fields of the Lagrangian
are replaced by renormalized parameters and fields, and a set of
counter terms:
\begin{eqnarray}
B_{\mu}  & \rightarrow & (Z_2^B)^{1/2} B_{\mu}  \nonumber \\
W_\mu^a  & \rightarrow & (Z_2^W)^{1/2} W_\mu^a  \nonumber \\
\psi_j^L & \rightarrow & (Z_L^j)^{1/2} \psi_j^L \nonumber \\
\psi_{j\sigma}^R & \rightarrow & (Z_R^{j\sigma})^{1/2} \psi_{j\sigma}^R
 \nonumber \\
H_i   & \rightarrow & Z_{H_i}^{1/2} H_i \nonumber \\
g_2   & \rightarrow & Z_1^W (Z_2^W)^{-3/2} g_2  \nonumber \\
g_1   & \rightarrow & Z_1^B (Z_2^B)^{-3/2} g_1  \nonumber \\
v_i   & \rightarrow & Z_{H_i}^{1/2} ( v_i - \delta v_i ) \nonumber \\
m_i^2 & \rightarrow & Z_{H_i}^{-1} ( m_i^2 + \delta m_i^2 ) \nonumber \\
m_{12}^2 & \rightarrow & Z_{H_1}^{-1/2} Z_{H_2}^{-1/2} ( m_{12}^2 + \delta
m_{12}^2 ) \ .
\label{glrenp}
\end{eqnarray}
The renormalization constants $Z_i$ are expanded by $Z_i \rightarrow
1 + \delta Z_i$.
This transforms the potential $V$
into a renormalized potential $V_{ren}$ and a counter term part $\delta V$:
\begin{equation}
 V \rightarrow V_{ren} (m_i^2, g_1, g_2, ... ) + \delta V \ .
\end{equation}
$V_{ren}$ is formally identical to (\ref{hpot}). Expressed in terms of the
rotated fields (\ref{rothig}) the coefficients of the quadratic part
\begin{equation}
 V_{ren} = \frac{1}{2} ( m_{h^0}^2 h^2 + m_{H^0}^2 H^2 + m_A^2 A^2 + ... ) \
, \end{equation}
are those of (\ref{baspara}, \ref{gl3}). The coefficients in the counter term
potential
\begin{equation}
 \delta V =  \delta t_{h^0} h + \delta t_{H^0} H + \frac{1}{2}
 ( \delta m_{h^0}^2 h^2 + \delta m_{H^0}^2 H^2 + \delta m_{A}^2 A^2 + ... )
\label{glpotct}
\end{equation}
are linear combinations of the counter terms in
(\ref{glrenp}).
\medskip \par
The on-shell renormalization conditions can be formulated in terms of
self energies, tadpoles and counter terms. In the following the
't Hooft-Feynman gauge is used and only the $g_{\mu \nu}$ components of
the vector boson propagators are considered.
The vector boson self energies
$\Sigma_{\gamma, Z, W, \gamma Z}$ are the transversal
components of the vector boson progagators $\Delta^V_{\mu \nu}$ ($V =
\gamma, Z, W$) \cite{bhs} : \\
\begin{eqnarray}
\Delta^V_{\mu \nu} (k) & = & - i g_{\mu \nu} \left( \frac{1}{k^2 - M_V^2}
- \frac{1}{k^2 - M_V^2} \ \Sigma_V (k^2) \ \frac{1}{k^2 - M_V^2} \right)
\nonumber \\
\Delta^{\gamma Z}_{\mu \nu} (k) & = & i g_{\mu \nu}  \frac{1}{k^2 - M_Z^2}
\ \Sigma_{\gamma Z} (k^2) \ \frac{1}{k^2} \ . \nonumber \\
\end{eqnarray}
The scalar boson self energies $\Sigma_S$ are related to the scalar
boson propagators $\Delta^S$ through: \\
\begin{equation}
\Delta^S (k)  =  i \left( \frac{1}{k^2 - M_S^2} - \frac{1}{k^2 - M_S^2}
\ \Sigma_S (k^2) \ \frac{1}{k^2 - M_S^2} \right) \ .
\end{equation} \\
The fermion self energies $\Sigma_f$ are defined by the fermion
propagator $S_f$ : \\
\begin{equation}
S_f (k) = \frac{i}{\not{k} - m_f} - \frac{i}{\not{k} - m_f} \ \Sigma_f (k) \
\frac{1}{\not{k} - m_f} \ .
\end{equation} \\
The $A^0$-$Z^0$ and $A^0$-$G^0$ mixing $\Sigma_{AZ}$, $\Sigma_{AG}$
are defined by the mixing propagator $\Delta^{A^0Z^0}$, $\Delta^{A^0G^0}$:
\begin{eqnarray}
 \Delta_\mu^{A^0Z^0} (k^2) &= & i k_\mu \ \frac{1}{k^2 - M_A^2 } \
 \Sigma_{AZ} (k^2) \ \frac{1}{k^2 - M_Z^2 } \nonumber \\
 \Delta^{A^0G^0} (k^2) &= &\frac{i}{k^2 - M_A^2 } \ \Sigma_{AG}
 (k^2) \ \frac{1}{k^2 - M_Z^2 } \ .
\end{eqnarray}
\medskip \par
In the following renormalized self energies are denoted by $\hat{\Sigma}$.
The
vector boson self energies, the $A^0$-boson self energy and the $A^0 Z^0$
mixing read:
\begin{eqnarray}
\hat{\Sigma}_Z (k^2) & = & {\Sigma}_Z (k^2) - \delta M_Z^2 + \delta Z_2^Z
( k^2 - M_Z^2 ) \nonumber \\
\hat{\Sigma}_W (k^2) & = & {\Sigma}_W (k^2) - \delta M_W^2 + \delta Z_2^W
( k^2 - M_W^2) \nonumber \\
\hat{\Sigma}_\gamma (k^2) & = & {\Sigma}_\gamma (k^2) + k^2 \delta
Z_2^\gamma   \nonumber \\
\hat{\Sigma}_{\gamma Z} (k^2) & = & {\Sigma}_{\gamma Z} (k^2) + M_Z^2
\frac{c_W}{s_W} \, ( \delta Z_1^\gamma - \delta Z_2^\gamma ) - k^2
\frac{c_W
s_W}{c_W^2 - s_W^2} ( \delta Z_2^Z - \delta Z_2^\gamma ) \nonumber \\[0.2cm]
\hat{\Sigma}_A (k^2) & = & \Sigma_A (k^2) - \delta m_A^2 + k^2
( \sin^2 \beta \, \delta Z_{H_1} + \cos^2 \beta \, \delta Z_{H_2} )
\nonumber \\
\hat{\Sigma}_{AZ} (k^2) & = & \Sigma_{AZ} (k^2) + M_Z \sin \beta
\cos \beta \ ( \delta Z_{H_1} - \delta Z_{H_2} ) \ .
\end{eqnarray}
The mass counter terms $\delta M_Z^2$, $\delta M_W^2$, $\delta m_A^2$
are defined in (\ref{glffff}) through the fundamental
renormalization constants in (\ref{glrenp}). They correspond to the
substitution:
\begin{eqnarray}
 M_Z^2 & \rightarrow & M_Z^2 + \delta M_Z^2 \nonumber \\
 M_W^2 & \rightarrow & M_W^2 + \delta M_W^2  \nonumber \\
 M_A^2 & \rightarrow & M_A^2 + \delta m_A^2 \ ,
\end{eqnarray}
where the unrenormalized masses are replaced by renormalized masses
and a mass counter term. The quantities
$$
s_W = \sin \theta_W \ , \ c_W = \cos \theta_W
$$
are the short-hand notations for the electroweak mixing angle in the
convention $c_W^2 = M_W^2/M_Z^2$ \cite{sirlin}.
\smallskip \par
The renormalized fermion self energy is given as follows:
\begin{eqnarray}
\hat{\Sigma}_f (k^2) & = & \not{k} \, \left( \Sigma_V^f (k^2) + \frac{\delta
Z_L + \delta Z_R^f}{2} \right)
+ \not{k} \gamma_5 \, \left( \Sigma_A^f (k^2) -  \frac{\delta Z_L - \delta
Z_R^f}{2} \right) \nonumber \\
& & + m_f \left( \Sigma_S^f (k^2) -  \frac{\delta Z_L + \delta Z_R^f}{2} -
\frac{\delta m_f}{m_f} \right) \ .
\label{glfermos}
\end{eqnarray}
\newpage
The renormalization
conditions for fixing the counter terms consist of:
\begin{itemize}
\item[1)]
the on-shell conditions in the gauge sector for $M_W$, $M_Z$, the
fermion masses $m_f$ and the
electromagnetic charge $e$, as in the minimal standard model.
They determine
the quantities $\delta g_1$, $\delta g_2$ and $\delta v^2 = \delta
( v_1^2 + v_2^2) $ due to (\ref{gmass}).
The on-shell conditions for the physical masses $M_Z$, $M_W$, $m_f$ are
defined as:
\begin{eqnarray}
 \Re e\  \hat{\Sigma}_Z (M_Z^2)       & = & 0  \nonumber \\
 \Re e\  \hat{\Sigma}_W (M_W^2)       & = & 0  \\
 \Re e\ \hat{\Sigma}_{f} (m_f^2) & = & 0 \ .
 \label{glfpbed}
\label{glrbed1}
\end{eqnarray}
\smallskip
The QED charge renormalization and residue conditions:
\begin{eqnarray}
 \hat{\Gamma}_\mu^{\gamma ee} ( k^2=0,\not{p}=\not{q}=m_e) & = & i e
\gamma_\mu
 \nonumber \\
 \Re e\ \frac{\partial}{\partial k^2} \hat{\Sigma}_\gamma ( k^2) |_{k^2=0}
 & = & 0  \nonumber \\
 \Re e\ \hat{\Sigma}_{\gamma Z} (0) & = & 0  \\
 \Re e\ \frac{1}{\not{k} - m_f} \hat{\Sigma}_{f} (k^2) |_{k^2=m_f^2} & = &
0 \ .
\label{glrbed2}
\end{eqnarray}
The residue condition for $f$ in (\ref{glrbed2}) is considered
for a fixed isospin component.
It can not be set for the upper and lower component of a fermion
doublet simultanously.
\item[2)] the tadpole conditions for vanishing renormalized tadpoles:\\[0.4cm]
\begin{figure}[h]
\epsfig{figure=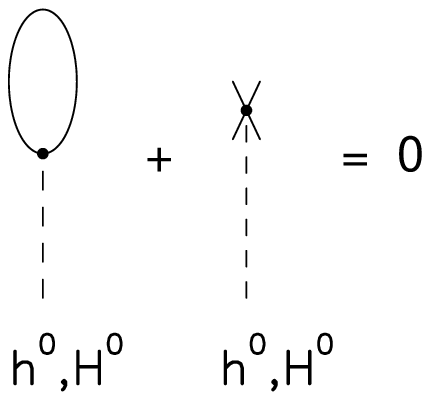,%
height=5.0cm,width=14.0cm,%
bbllx=0pt,bblly=520pt,bburx=405pt,bbury=751pt}
\end{figure}
\vspace{-5.3cm}
$$
\hspace*{-4.5cm} T_{H^0 (h^0)}+ \delta t_{H^0 (h^0)} = 0
$$ \\[3.1cm]
They ensure that $v_1, v_2$ determine the minimum of the one-loop potential.\\
\item[3)]
the on-shell and residue condition for the $A^0$-boson: \\
\begin{equation}
 \Re e\ \hat{\Sigma}_{A} (M_{A}^2) = 0 \\
\label{glpabd}
\end{equation}
\begin{equation}
 \Re e\ \frac{\partial}{\partial k^2} \hat{\Sigma}_{A^0} ( k^2 )  \mid _{
 k^2 = M_A^2} = 0 \ .
\label{glrbed3}
\end{equation}
By (\ref{glpabd}) $\delta m_A^2$ is determined in terms of the
on-shell self energy of $A^0$. \\
\item[4)]
renormalization of $\tan \beta$: \\
The relation $\tan \beta = v_2 / v_1$ in terms of the " true vacua "
is maintained by the condition
\begin{equation}
 \frac{ \delta v_1}{v_1} = \frac{ \delta v_2}{v_2} \ .
\end{equation}
\item[5)]
the vanishing of the $A$-$Z$ mixing on-shell: \\
\begin{eqnarray}
 \Re e \ \hat{\Sigma}_{A^0Z^0} (M_A^2) & = & 0 \ .
\label{glrba}
\end{eqnarray} \\
\end{itemize}
By this set of conditions, the extra input for the Higgs sector
(besides the parameters already used for the gauge sector) is given
by $\tan \beta$ and the physical mass of the $A^0$-boson.
\medskip \par
The mass counter terms of the $W^\pm$, $Z^0$ and $A^0$ bosons in
the on-shell condition (\ref{glrbed1}) are
\begin{eqnarray}
\Re e\  \Sigma_Z (M_Z^2) & = & \delta M_Z^2 \nonumber \\
 & = & M_Z^2 \left( 2 \delta Z_1^Z - 3 \delta Z_2^Z + \cos^2 \beta \
 \delta Z_{H_1} + \sin^2 \beta \ \delta Z_{H_2} - 2  \cos^2 \beta \ \frac
 {\delta v_1}{v_1} \right. \nonumber \\ &  & \left.
 - 2 \sin^2 \beta \ \frac{\delta v_2}{v_2} \right) \nonumber
 \\
 \Re e\  \Sigma_W (M_W^2) & = & \delta M_W^2    \nonumber \\
 & = & M_W^2 \left( -\frac{c_W^2}{c_W^2 - s_W^2} ( 3 \delta Z_2^Z - 2
 \delta Z_1^Z ) + \frac{s_W^2}{c_W^2 - s_W^2} ( 3 \delta Z_2^\gamma - 2
 \delta Z_1^\gamma )  \right. \nonumber \\ & &
\left.  + \cos^2 \beta \ \delta Z_{H_1}
 + \sin^2 \beta  \ \delta  Z_{H_2}  - 2  \cos^2 \beta \
\frac{\delta v_1}{v_1} - 2 \sin^2 \beta \ \frac{\delta v_2}{v_2} \right)
\nonumber \\
\Re e\  \Sigma_A (M_A^2) & = & \delta m_{A}^2 -
M_A^2 \ ( \sin^2 \beta \ \delta Z_{H_1} + \cos^2 \beta \ \delta Z_{H_2} )
\nonumber \\ & = &
\frac{1}{2} \ ( \ \sin^2 \beta \ \delta m_1^2 +
\cos^2 \beta \ \delta m_2^2 - \sin 2 \beta
\ \delta m_{12}^2 \ )  - \frac{M_Z^2}{4} \ \cos^2 2 \beta
\nonumber \\ & & ( \ \delta Z_Z - 2
\frac{\delta v}{v} + \delta Z_H \ ) -
M_A^2 \ ( \sin^2 \beta \ \delta Z_{H_1} + \cos^2 \beta \ \delta Z_{H_2} )
 \ , \nonumber \\
\label{glffff}
\end{eqnarray}
where $\delta v /v = \delta v_i / v_i$, $\delta Z_Z = 2 \delta Z_1^Z -
3 \delta Z_2^Z$ and $ \delta Z_H = \delta Z_{H_1} + \delta Z_{H_2}$.
The photon and $Z^0$-boson vertex and field renormalization constants
are as in the minimal standard model:
\begin{eqnarray}
 \delta Z_1^Z & = & c_W^2 \delta Z_1^W + s_W^2 \delta Z_1^B \nonumber
 \\[0.2cm]
 & = & -\Sigma_\gamma' (0) - \frac{ 3 c_W^2 - 2 s_W^2}{ s_W c_W }
 \frac{ \Sigma_{\gamma Z} (0)}{M_Z^2} + \frac{c_W^2 - s_W^2}{s_W^2}
 \left( \frac{ \Sigma_Z (M_Z^2)}{M_Z^2} - \frac{ \Sigma_W (M_W^2)}{M_W^2}
 \right)  \nonumber \\[0.2cm]
 \delta Z_2^Z & =  & c_W^2 \delta Z_2^W + s_W^2 \delta Z_2^B \nonumber
 \\[0.2cm]
 & = & -\Sigma_\gamma' (0) - 2 \frac{ c_W^2 - s_W^2}{ s_W c_W }
 \frac{ \Sigma_{\gamma Z} (0)}{M_Z^2} + \frac{c_W^2 - s_W^2}{s_W^2}
 \left( \frac{ \Sigma_Z (M_Z^2)}{M_Z^2} - \frac{ \Sigma_W (M_W^2)}{M_W^2}
 \right) \nonumber \\[0.2cm]
 \delta Z_1^\gamma & = & -\Sigma_\gamma' (0) -\frac{s_W}{c_W} \frac{
 \Sigma_{\gamma Z} (0)}{M_Z^2} \nonumber \\[0.2cm]
 \delta Z_2^\gamma & = & -\Sigma_\gamma' (0) \ .
\end{eqnarray}
The Higgs field renormalization constants $\delta Z_{H_i}$ are obtained
from (\ref{glrbed3}, \ref{glrba}):
\begin{eqnarray}
 \delta Z_{H_1} & = & -\Sigma_A'(M_A^2) - \frac{ \cot \beta}{M_Z}
   \ \Sigma_{A Z} (M_A^2)  \nonumber \\
 \delta Z_{H_2} & = & -\Sigma_A'(M_A^2) + \frac{ \tan \beta}{M_Z}
   \ \Sigma_{A Z} (M_A^2) \ .
\end{eqnarray}
{}From the Higgs potential counter term $\delta V$ (\ref{glpotct}) one
obtaines the tadpole counter terms expressed by:
\begin{eqnarray}
 \delta t_{h^0} & = & - \sqrt{2} v_1 \sin \alpha \ \delta m_1^2 +
 \sqrt{2} v_2 \cos \alpha \ \delta m_2^2 + \sqrt{2} ( v_1 \cos
 \alpha - v_2 \sin \alpha ) \ \delta m_{12}^2 + R_{h^0}  \nonumber \\
 \delta t_{H^0} & = & \sqrt{2} v_1 \cos \alpha \ \delta m_1^2 +
 \sqrt{2} v_2 \sin \alpha \ \delta m_2^2 + \sqrt{2} ( v_1 \sin \alpha
 + v_2 \cos \alpha ) \ \delta m_{12}^2 + R_{H^0} \nonumber \ .  \\
\end{eqnarray}
The quantities $R_{h^0}$, $R_{H^0}$, $R_A$ contain
the gauge sector counter terms in the Higgs self interaction part:
\begin{eqnarray}
R_{h^0} & = &  m_1^2 \sin \alpha \, \delta v_1 -
m_2^2 \cos \alpha \, \delta v_2 +
m_{12}^2 ( \sin \alpha \, \delta v_2 - \cos \alpha \, \delta v_1 )
\nonumber \\ & & + \frac{ M_W M_Z^2}{g_2} \left( \
 -\cos 2 \beta \sin (\beta + \alpha)
( \delta Z_Z - 3 \delta v /v ) \right. \nonumber \\ & & \left.
 -\frac{1}{2} \delta Z_{H} \sin 2 \beta \cos (\beta + \alpha) -
 2 \delta Z_{H_1} \cos^3 \beta \sin \alpha + 2 \delta Z_{H_2}
 \sin^3 \beta \cos \alpha \right)  \nonumber \\[0.3cm]
R_{H^0} & = & -m_1^2 \cos \alpha \, \delta v_1 -
m_2^2 \sin \alpha \, \delta v_2 -  m_{12}^2
 ( \cos \alpha \, \delta v_2 + \sin \alpha \, \delta v_1 )
 \nonumber \\ & &
 + \frac{ M_W M_Z^2}{g_2}  \left( \
 \cos 2 \beta \cos (\beta + \alpha)
 ( \delta Z_Z - 3 \delta v /v ) \right. \nonumber \\ & & \left.
 -\frac{1}{2} \delta Z_{H} \sin 2 \beta \sin (\beta + \alpha)
 + 2 \delta Z_{H_1} \cos^3 \beta \cos \alpha + 2 \delta Z_{H_2}
 sin^3 \beta \sin \alpha \right) \nonumber \\[0.3cm]
R_A & = & - \frac{M_Z^2}{4} \ \cos^2 2 \beta \ (  \ \delta Z_Z - 2
\frac{\delta v}{v} + \delta Z_H \ ) \ . \nonumber
\end{eqnarray}
Introducing the linear combinations:
\begin{eqnarray}
\delta X & = & \frac{1}{v} \ \left( \ \cos (\beta -\alpha) \
( \delta t_{H^0} - R_{H^0} ) + \sin (\beta -\alpha) \
( \delta t_{h^0} - R_{h^0} ) \ \right)    \nonumber \\
\delta Y & = & \frac{1}{v} \ \left( \ \sin (\beta -\alpha) \
( \delta t_{H^0} - R_{H^0} ) - \cos (\beta -\alpha) \
( \delta t_{h^0} - R_{h^0} ) \ \right) \ ,
\end{eqnarray}
where $ v^2 = v_1^2 + v_2^2$, yields the mass counter terms
$\delta m_i^2$, $\delta m_{12}^2$
\begin{eqnarray}
\delta m_1^2 & = & \delta X  \ \cos^2 \beta
+ \delta Y \ \sin 2 \beta + 2 \sin^2 \beta \ (\delta m_A^2 - R_A)
\nonumber
\\ \delta m_2^2 & = & \delta X \ \sin^2 \beta -
\delta Y \ \sin 2 \beta + 2 \cos^2 \beta \ (\delta m_A^2 - R_A)
\nonumber \\ \delta m_{12}^2 & = & \frac{1}{2} \sin 2 \beta \ \delta X -
\cos 2 \beta \ \delta Y - \sin 2 \beta \ (\delta m_A^2 - R_A) \ .
\end{eqnarray}
The complete set of gauge and Higgs counter terms is now fixed by the
self energies and tadpoles. This subset is required for
 the calculation of the $h^0$, $H^0$ and $h^0 H^0$
propagators at the one-loop level in the next chapter.
\par \smallskip
The fermion mass counter term and field renormalization
constant follow from (\ref{glfpbed}, \ref{glrbed2}):
\begin{eqnarray}
 \Re e\  \Sigma_f(m_f^2) & = & \delta m_f^2 + \frac{m_f}{2} ( \delta Z_L^f +
 \delta  Z_R^f ) - \not{k} ( \delta Z_V^f - \gamma_5 \delta Z_A^f )
|_{\not{k} = m_f} \nonumber \\
 \frac{\delta m_f^2}{m_f^2}
 & = & \Sigma_V^f (m_f^2) +  \Sigma_S^f (m_f^2) \nonumber \\
 \delta Z_V^f & =& \frac{1}{2} ( \delta Z_L^f + \delta Z_R^f ) \nonumber \\
 & = & - \Sigma_V^f (m_f^2) - 2 m_f^2 ( \Sigma_V^{'f} (m_f^2) +
 \Sigma_S^{'f} (m_f^2) ) \nonumber \\
 \delta Z_A^f &=& \frac{1}{2} ( \delta Z_L^f - \delta Z_R^f ) \nonumber \\
 & = &  \Sigma_A^f (m_f^2) \ ,
\label{glfrec}
\end{eqnarray}
where the decomposition of $ \Sigma_f $ into the
invariant functions $\Sigma_{V,A,S}^f$ is applied:
\begin{equation}
\Sigma_f (k) = \not{k} \Sigma_V^f (k^2) + \not{k} \gamma_5
               \Sigma_A^f (k^2) + m_f \Sigma_S^f (k^2) \ .
\end{equation}
\smallskip \par
%
%
\subsection{Physical neutral Higgs masses in the MSSM}
\setcounter{equation}{0}\setcounter{footnote}{0}
\vspace*{0.1cm} \hspace*{0.5cm}
Radiative corrections to the neutral scalar Higgs particles require
the calculation of the renormalized Higgs self energies and the
$h^0$-$H^0$ mixing. They are obtained by summing the loop
diagrams and the counter terms:
\begin{eqnarray}
 \hat{\Sigma}_{h^0} (k^2) & = & \Sigma_{h^0} (k^2) + k^2 \
 ( \delta Z_{H_1} \sin^2 \alpha + \delta Z_{H_2} \cos^2 \alpha )
 - \delta m^2_{h^0} \nonumber \\[0.2cm]
 \hat{\Sigma}_{H^0} (k^2) & = & \Sigma_{H^0} (k^2) + k^2 \
 ( \delta Z_{H_1} \cos^2 \alpha + \delta Z_{H_2} \sin^2 \alpha )
 - \delta m^2_{H^0} \nonumber \\[0.2cm]
 \hat{\Sigma}_{H^0 h^0} (k^2) & = & \Sigma_{H^0 h^0} (k^2) + k^2 \
 \sin \alpha \cos \alpha \ ( \delta Z_{H_2} - \delta Z_{H_1} )
 - \delta m^2_{H^0 h^0} \ .
\label{selfh}
\end{eqnarray}
The mass counter terms $\delta m^2_{h^0}$, $\delta m^2_{H^0}$
and $\delta m^2_{H^0 h^0}$ are completely fixed in the Higgs potential
counter terms: \\
\begin{eqnarray*} \lefteqn{
\delta m_{h^0}^2  =  \cos^2 (\beta - \alpha) \ \delta m_{A}^2
 \ + \ \frac{g_2 \sin^2 (\beta - \alpha) \cos (\beta - \alpha) }
 {2 M_W} \ T_{H^0} \ + \ \sin^2 (\beta + \alpha) \ \Sigma_Z (M_Z^2) }
 \nonumber  \\[0.1cm] & &
 \ - \frac{g_2 \sin (\beta - \alpha) (
 1 +   \cos^2 (\beta - \alpha) ) }{2 M_W} \ T_{h^0} \
 - \ M_Z^2 \sin (\beta + \alpha)   \sin (\beta - \alpha) \ ( \delta Z_{H_1}
 -  \delta Z_{H_2} )
 \nonumber \\[0.2cm] & & + \ M_Z^2 \sin^2 (\beta + \alpha) \
 (\sin^2 \beta \
 \delta Z_{H_1} + \cos^2 \beta \ \delta Z_{H_2} )
 \nonumber \\[0.5cm]   \lefteqn{
\delta m_{H^0}^2  =  \sin^2 (\beta - \alpha) \ \delta m_{A}^2 \
 - \ \frac{g_2 \cos (\beta - \alpha) ( 1 + \sin^2 (\beta - \alpha) ) }
 {2 M_W} \ T_{H^0} } \nonumber \\[0.2cm] & &
 + \ \cos^2 (\beta + \alpha)
  \Sigma_Z (M_Z^2) + \ \frac{g_2 \cos^2 (\beta - \alpha)
  \sin (\beta - \alpha) }{2 M_W} \ T_{h^0}
 \nonumber \\[0.2cm] & &  + \
  M_Z^2 \cos (\beta + \alpha)   \cos (\beta - \alpha)
 ( \delta Z_{H_1} - \delta Z_{H_2} )
 \nonumber \\[0.2cm] & &
 + \ M_Z^2 \cos^2 (\beta + \alpha)
 ( \sin^2 \beta \
 \delta Z_{H_1} + \cos^2 \beta \ \delta Z_{H_2} )
  \nonumber \\[0.5cm]   \lefteqn{
\delta m_{H^0 h^0}^2  =  - \sin (\beta - \alpha) \cos (\beta - \alpha)\
 \delta m_A^2
 \ + \ \frac{g_2 \sin^3 (\beta - \alpha) }
 {2 M_W} \ T_{H^0} \ + \ \frac{g_2 \cos^3 (\beta - \alpha) }{2 M_W} \
 T_{h^0} }  \nonumber \\[0.2cm] & &
 - \ \cos (\beta + \alpha) \sin (\beta + \alpha) \
 \Sigma_Z (M_Z^2) \ - \ M_Z^2 \sin \alpha
 \cos \alpha \ ( \delta Z_{H_1} -  \delta Z_{H_2} ) \nonumber \\[0.2cm] &
 & - \ M_Z^2 \cos (\beta + \alpha)
 \sin (\beta + \alpha) ( \sin^2 \beta \ \delta Z_{H_1} + \ \cos^2 \beta \
 \delta Z_{H_2} ) \ .
\end{eqnarray*} \\
\par
The propagator matrix of the scalar neutral Higgs particles is
diagonal in lowest order perturbation theory. Quantum effects
give rise to mixing between the light and heavy Higgs boson states.
Therefore the $(h^0,H^0)$ propagator matrix in one loop order is a
$2 \times 2$ matrix. The inverse of the propagator matrix is
\begin{equation}
 \Delta^{-1} = -i \ \left ( \begin{array}{cc}
  k^2 - m^2_{h^0} + \hat{\Sigma}_{h^0} (k^2) &
  \hat{\Sigma}_{h^0H^0} (k^2) \\
  \hat{\Sigma}_{h^0H^0} (k^2) &
  k^2 - m^2_{H^0} + \hat{\Sigma}_{H^0} (k^2)  \\
  \end{array}   \right ) \ ,
\label{glbas1}
\end{equation}
where $\hat{\Sigma}_{h^0}, \hat{\Sigma}_{H^0}, \hat{\Sigma}_{h^0H^0} $
are the renormalized self energies and mixing
as indicated in (\ref{selfh}).
The entries in the propagator matrix
\begin{equation}
 \Delta = i \left ( \begin{array}{ll}   \Delta_{h^0} &  \Delta_{h^0H^0}
  \\
  \Delta_{h^0H^0} & \Delta_{H^0}  \\
  \end{array}   \right ) \ ,
\label{glhhmix}
\end{equation}
are the diagonal and non-diagonal propagators:
\begin{eqnarray}
 \Delta_{h^0} & = & \frac{1}{k^2 - m^2_{h^0} + \hat{\Sigma}_{h^0} (k^2)
  - \frac{\hat{\Sigma}_{h^0H^0}^2 (k^2) }
  { k^2 - m^2_{H^0} + \hat{\Sigma}_{H^0} (k^2) }} \nonumber \\
 \Delta_{H^0} & = & \frac{1}{k^2 - m^2_{H^0} + \hat{\Sigma}_{H^0} (k^2)
  - \frac{\hat{\Sigma}_{h^0H^0}^2 (k^2) }
  { k^2 - m^2_{h^0} + \hat{\Sigma}_{h^0} (k^2) }} \nonumber \\
 \Delta_{h^0H^0} & = & \frac{ - \ \hat{\Sigma}_{h^0H^0} (k^2) }
  { (k^2 - m^2_{h^0} + \hat{\Sigma}_{h^0} (k^2) )
    (k^2 - m^2_{H^0} + \hat{\Sigma}_{H^0} (k^2) ) - \hat{\Sigma}_{h^0H^0}^2
  (k^2) } \ .
\label{gl7}
\end{eqnarray}
The physical one-loop masses of the scalar neutral Higgs particles
are the real parts of the one-loop propagator matrix poles. The poles of
the propagator system are the solutions of the equation
\begin{equation}
 ( k^2 - m^2_{h^0} + \hat{\Sigma}_{h^0} (k^2) \ )
 ( k^2 - m^2_{H^0} + \hat{\Sigma}_{H^0} (k^2) \ ) - ( \hat{\Sigma}_{h^0 H^0}
 (k^2)\ )^2  = 0 \ .
\label{gl8}
\end{equation}
The physical Higgs masses are denoted by capital $M_{h^0}, M_{H^0}$ in
order to distinguish them from the formal tree level Higgs masses
$m_{h^0}, m_{H^0}$ in Eq. (\ref{gl3}).
\par \smallskip
In the numerical analysis the Fermi constant $G_F$ is taken as the input
parameter from $\mu^-$ decay. $G_f$ is related to $M_W$ by:
\begin{equation}
 G_F = \frac{\pi \alpha}{\sqrt{2} s_W^2 M_W^2} \cdot \frac{1}{1 -
\Delta r } \ ,
\end{equation}
where $\Delta r $ is the full MSSM radiative correction to the
$\mu^-$ decay amplitude \cite{sola}.
\par
\smallskip
In the following we discuss the effects from virtual supersymmetric
particles in radiative corrections to the physical light and
heavy Higgs mass.
\subsection{Discussion}
\setcounter{equation}{0}\setcounter{footnote}{0}
\medskip
$\bullet$ Higgs mass approximation formulae $\cal{O} \rm (m_t^4)$
\medskip \par
The complete numerical analysis of the physical neutral Higgs masses
confirms that virtual fermions and sfermions give the
largest contributions to radiative corrections through the Yukawa couplings.
The top-stop loops give rise to the dominant term increasing
with the top quark mass $\sim m_t^4$.
These leading contributions to the renormalized self energies
$\hat{\Sigma}_{h^0}, \hat{\Sigma}_{H^0}$, $\hat{\Sigma}_{h^0H^0}$ are
given by
\begin{eqnarray}
\hat{\Sigma}_{h^0} (k^2) & = & - \omega_t \, \cos^2 \alpha \nonumber \\
\hat{\Sigma}_{H^0} (k^2) & = & - \omega_t \, \sin^2 \alpha  \nonumber \\
\hat{\Sigma}_{h^0H^0} (k^2) & = & - \omega_t \, \sin \alpha \cos \alpha \ ,
\label{glltop}
\end{eqnarray}
where
\begin{equation}
 \omega_t = \frac{N_C G_F m_{t}^4 }{\sqrt{2}  \pi^2  \sin^2 \beta} \
\log \ ( \frac{m_{\tilde{t}_L} m_{\tilde{t}_R} }{m_{t}^2} ) \ . \\
\label{leadom}
\end{equation}
Thereby the masses of the left and right stop squark states are
$m_{\tilde{t}_L}$, $m_{\tilde{t}_R}$ and
$N_C$ is the number of colors. In this approximation, the
solution of equation (\ref{gl8}) is the following simple modification
of Eq. (\ref{gl3}):
\begin{eqnarray}
 M^2_{H,h,\, eff} & = & \frac{M_A^2 + M_Z^2 + \omega_t}{2} \nonumber \\
 & & \pm \, \sqrt{
  \frac{ (M_A^2 + M_Z^2)^2 + \omega^2_t}{4} - M_A^2 M_Z^2 \cos^2 2\beta +
  \frac{\omega_t \cos 2\beta}{2}  (M_A^2 - M_Z^2) } \ .
\label{glmapp} \nonumber \\
\end{eqnarray}
As compared to the lowest order behaviour, the following
striking changes occur: \\
\begin{tabular}[t]{ll}
- & \begin{minipage}[t]{15.0cm}
 the upper limit for the light Higgs mass $M_{h^0}$ exceeds the tree
 level limit $m_{h^0} \leq M_Z$ significantly. \end{minipage} \\
- & \begin{minipage}[t]{15.0cm}
 the spectrum is no longer symmetric under $\tan \beta \leftrightarrow
 1/\tan \beta$.   \end{minipage} \\
- & \begin{minipage}[t]{15.0cm}
 $M_{h^0} \neq 0$ for $M_A \rightarrow 0$ in general.
\end{minipage} \\
\end{tabular}\\[0.2cm]
These features are already described by the approximate formulae
(\ref{glmapp}).
In the next sections the complete one-loop calculation is
compared with the approximation formulae (\ref{glmapp}).\\[0.3cm]
$\bullet$ Higgs mass dependence on $M_A$, $\tan \beta$
\medskip \par
Figs. 1a,b show the light and heavy physical Higgs masses $M_h$, $M_H$
as functions of the input parameters $M_A$, $\tan \beta$.
The values for $M_A$, $\tan \beta$ are varied in the range
$$
0.5 \le \tan \beta \le m_t/m_b
$$
$$
0 \le M_A \le 300 \ \mbox{GeV} \ .
$$
For the top quark mass we take the value $m_t = 175$ GeV in accordance
with the experimental indication
$m_t = 174.1 \pm 17 $ GeV \cite{top}. The SUSY soft breaking
parameters are chosen at
$\mu = 100$ GeV, $M = 400$ GeV, $m_{sf} = 500$ GeV. $m_{sf}$ defines
the sfermion soft breaking parameter, which is assumed to be equal
for all squarks and sleptons (\ref{sfmatrix}).
A diagonal squark and slepton mass matrix is also assumed.
The complete one-loop result is displayed in Figs. 1a,b. The differences
with the approximation formulae (\ref{glmapp}), are
depicted in Tab. \ref{tab1}, for $M_A = 300$ GeV:
\begin{table}[ht]
\begin{center}
\begin{tabular}{l||r|r||r|r}
 $\tan \beta$ & $M_{h,\, eff}$ [GeV] & $M_{h,\, 1-loop}$ [GeV] &
 $M_{H,\, eff}$ [GeV] & $M_{H,\, 1-loop}$ [GeV]  \\
 \hline & & & & \\
 0.5  &  87.4   &  92.4   &  340.0   &  352.0     \\
 2    &  86.5   &  83.0   &  311.5   &  312.2     \\
 5    & 108.9   & 105.8   &  302.7   &  302.9     \\
 10   & 113.5   & 110.7   &  300.7   &  300.8     \\
 30   & 115.0   & 113.7   &  300.1   &  300.1     \\
\end{tabular}
\end{center}
\caption{Physical neutral Higgs masses in the
approximation of Eq. (4.3) $M_{h,\, eff}$, $M_{H,\, eff}$
and in the full one-loop result $M_{h,\, 1-loop}$, $M_{H,\, 1-loop}$.}
\label{tab1}
\end{table}
\par
A nearly constant light Higgs mass plateau is obtained for fixed
$\tan \beta$ values and pseudoscalar masses $M_A > 200$ GeV (Fig. 1a).
In the full one-loop result, lower $\tan \beta$ values yield
larger light Higgs masses  $M_h$ ($5$ GeV for $\tan \beta = 0.5$) compared to
the approximate result.
The exact value can be smaller than the approximate result ($\approx
1-3$ GeV for $\tan \beta \geq 2$).\\[0.3cm]
$\bullet$ Higgs mass dependence on $m_t$
\smallskip \par
Figs. 1c,d illustrate the expected top quark dependence of the light
and heavy Higgs masses. SUSY soft breaking parameters and $\tan \beta$ values
are taken from Figs. 1a,b and a pseudoscalar mass $M_A = 200$ GeV is
selected. The leading $\sim m_t^4$ dependence in (\ref{leadom}) is
shown for a top quark mass in the experimental $2 \sigma$ top mass range
between $150$ GeV $< m_t <$ $200$ GeV. In case of the light Higgs this
$m_t$ dependence change the Higgs mass by $\pm 12$ GeV from its mean
value at $m_t = 175$ GeV and $M_A = 200$ GeV.
The heavy Higgs mass shows sizeable $m_t$
dependent effects only for low $\tan \beta$ values. For $\tan \beta > 5$
the heavy Higgs mass deviates by $\pm 1$ GeV within the $2 \sigma$ top mass
range.
The quality of the approximation formulae at $m_t = 150$ GeV compared
to the full one-loop result is shown in Tab. \ref{tab2} where the
result of Eq. (\ref{glmapp}) is listed.
\begin{table}[ht]
\begin{center}
\begin{tabular}{l||r|r||r|r}
 $\tan \beta$ & $M_{h,\, eff}$ [GeV] & $M_{h,\, 1-loop}$ [GeV] &
 $M_{H,\, eff}$ [GeV] & $M_{H,\, 1-loop}$ [GeV]  \\
 \hline & & & & \\
 0.5  &  77.1   &   78.4     &  240.8   &   243.6    \\
 2    &  72.6   &   68.4     &  216.4   &   216.7    \\
 5    &  98.4   &   95.2     &  204.3   &   204.5    \\
 10   & 104.3   &  101.3     &  201.2   &   201.3    \\
 30   & 106.2   &  104.1     &  200.1   &   200.2    \\
\end{tabular}
\end{center}
\caption{Physical neutral Higgs masses in the
approximation of Eq. (4.3) $M_{h,\, eff}$, $M_{H,\, eff}$
and in the full one-loop result $M_{h,\, 1-loop}$, $M_{H,\, 1-loop}$.}
\label{tab2}
\end{table}
\par
The results for $\tan \beta \geq 2 $ in the full calculation are $\approx
2-4$ GeV smaller than the approximation of Eq. (\ref{glmapp}). A value
$\tan \beta = 0.5$ yields a difference of $1$ GeV between the
full result and the approximation formulae.\\[0.3cm]
$\bullet$ Higgs mass dependence on squarks and sleptons
\smallskip \par
The squark and slepton sector of the MSSM is described by a
$2 \times 2$ mass matrix:
\begin{equation}
 \cal M_{\rm \tilde{f}}^{\rm 2} \rm =
 \left( \begin{array}{cc} \tilde{M}^2_{\tilde{Q}} + M_Z^2 \cos 2 \beta
 ( I_3 - Q_f s_W^2 ) + m_f^2 & m_f ( A_f + \mu \{\cot , \tan \} \beta ) \\
 m_f ( A_f + \mu \{ \cot , \tan \} \beta ) & \tilde{M}^2_{\tilde{U},
 \tilde{D}} + M_Z^2 \cos 2 \beta Q_f s_W^2 + m_f^2 \end{array} \right) \ , \\
\label{sfmatrix}
\end{equation}
with SUSY soft breaking parameters $\tilde{M}_{\tilde{Q}}$,
$\tilde{M}_{\tilde{U}, \tilde{D}}$, $A_f$, and $\mu$.
In the following discussion the soft breaking parameters are taken to
be equal $m_{sf} = \tilde{M}_{\tilde{Q}} = \tilde{M}_{\tilde{U}, \tilde{D}}$.
Up and down type sfermions in (\ref{sfmatrix}) are distinguished by
setting f=u,d and the $\{u,d\}$ entries in the parenthesis.
The parameter $\mu$ in the off-diagonal matrix elements in (\ref{sfmatrix})
is also present in the gaugino sector. The sfermion masses, obtained from
diagonalizing (\ref{sfmatrix}) are:
\begin{equation}
 m_{\tilde{f}_i}^2 = \frac{1}{2} (\rm Tr \cal M_{\rm \tilde{f}}^{\rm 2} \rm
 \pm \sqrt{ ( Tr \cal M_{\rm \tilde{f}}^{\rm 2} \rm ) ^2 - 4
 Det \cal M_{\rm \tilde{f}}^{\rm 2} \rm } \ ) \ , \ i=1,2 \ ,
\end{equation}
where the corresponding rotation matrices are described by the sfermion
mixing angle $\tilde{\theta}_f$:
\begin{equation}
 \tan 2 \tilde{\theta}_f = \frac{2 m_f ( A_f - \mu \{ \cot \beta,
 \tan \beta \} ) }{ M_{\tilde{f}}^2 - M_{\{\tilde{U},\tilde{D}\}}^2
 + M_Z^2 (I_3 - 2 Q_q s_W^2) \cos 2 \beta } \ .
\end{equation}
As a first step, a diagonal mass matrix is obtained by setting:
\begin{equation}
A_f + \mu  \{\cot , \tan \} \beta  = 0 \ .
\label{sfbed}
\end{equation}
Together with the present experimental lower bounds on squark masses
\cite{pdg},
Figs. 2a,b show the light and heavy Higgs masses in the range
$100$ GeV $< m_{sf} <$ $1$ TeV, for the parameters of Figs. 1a,b
and $M_A = 200$ GeV.
The result of Eq. (\ref{glmapp}) is listed in Tab. \ref{tab3} for
$m_{sf} = 200$
GeV ($1$ TeV) and the parameters of Figs. 2a,b.
\begin{table}[ht]
\begin{center}
\begin{tabular}{l||r|r||r|r}
 $\tan \beta$ & $M_{h,\, eff}$ [GeV] & $M_{h,\, 1-loop}$ [GeV]
  & $M_{H,\, eff}$ [GeV] &  $M_{H,\, 1-loop}$ [GeV]  \\
 \hline & & & & \\
 & ($m_{sf} = 200$ GeV) & ($m_{sf} = 200$ GeV) & ($m_{sf} = 200$ GeV) &
   ($m_{sf} = 200$ GeV) \\
 0.5  &  70.1   &  69.6   & 230.2   &  233.4  \\
 2    &  65.2   &  61.2   & 215.3   &  215.7  \\
 5    &  92.5   &  89.9   & 204.1   &  204.2  \\
 10   &  98.5   &  96.1   & 201.1   &  201.2  \\
 30   & 100.4   & 106.0   & 200.1   &  200.2  \\
 \hline & & & & \\
 & ($m_{sf} = 1$ TeV) & ($m_{sf} = 1$ TeV) & ($m_{sf} = 1$ TeV) &
   ($m_{sf} = 1$ TeV) \\
 0.5  &  94.1   &  96.6   &  281.0  & 287.4   \\
 2    &  96.1   &  92.5   &  221.3  & 221.3   \\
 5    & 119.3   & 115.6   &  205.7  & 205.7   \\
 10   & 125.1   & 121.4   &  201.6  & 201.7   \\
 30   & 127.1   & 123.6   &  200.2  & 200.3   \\
\end{tabular}
\end{center}
\caption{Physical neutral Higgs masses in the
approximation of Eq. (4.3) $M_{h,\, eff}$, $M_{H,\, eff}$
and in the full one-loop result $M_{h,\, 1-loop}$, $M_{H,\, 1-loop}$.}
\label{tab3}
\end{table}
\par
A squark soft breaking parameter $m_{sf} = 200$ GeV ($1$ TeV) yields
a result in the full one-loop calculation which is by $3$-$6$ ($3$-$4$) GeV
smaller than the formulae (\ref{glmapp}) with  $\tan \beta \geq 2$.
\par \smallskip
A non-diagonal mass matrix (\ref{sfmatrix}) induces a mixing of the
left and right sfermion states described by the sfermion
mixing angle $\tilde{\theta}_f$.
As a consequence, the Higgs couplings to sfermions contain the
mixing angles $\tilde{\theta}_f$ and modify the approximation
formulae (\ref{glmapp}) sizeably. The leading $\sim m_t^4$
contribution to the physical light Higgs mass is obtained in
replacing (\ref{leadom}) by
\begin{eqnarray}
 \omega_t = \frac{N_C G_F m_{t}^4 }{\sqrt{2} \pi^2 \sin^2 \beta} \
  & & \hspace{-0.5cm}
 \left(  \log \ ( \frac{m_{\tilde{t}_1} m_{\tilde{t}_2} }{m_{t}^2} )
+ \frac{A_t ( A_t + \mu \cot \beta)}{ m_{\tilde{t}_1}^2 -
 m_{\tilde{t}_2}^2 } \log \frac{ m_{\tilde{t}_1}^2 }{ m_{\tilde{t}_2}^2 }
 \right. \nonumber \\ & &  \hspace*{-0.5cm}    \left. +
 \frac{A_t^2 ( A_t + \mu \cot \beta)^2 }{ ( m_{\tilde{t}_1}^2 -
 m_{\tilde{t}_2}^2 )^2} \left( 1 - \frac{m_{\tilde{t}_1}^2 + m_{\tilde{t}_2}^2}
{m_{\tilde{t}_1}^2 - m_{\tilde{t}_2}^2} \log \frac{m_{\tilde{t}_1}}
{m_{\tilde{t}_2}} \right) \ \right) \ . \nonumber \\
\label{leadom2}
\end{eqnarray}
In Figs. 2c,d the light Higgs mass is plotted as a function of the
parameters $\mu$ and $A = A_u = A_d$ for $\tan \beta = 30 (2)$.
A large $\tan \beta = 30$ in
Fig 2c shows a variation of the light Higgs by $2$ GeV for $\mu$ in the range
$-250$ GeV  $< \mu <$  $250$ GeV.
Small $\tan \beta = 2$ values in Fig. 2d give
strong deviations of the light Higgs mass with $\mu$. Not all
squark and gaugino masses are experimentally allowed in Fig. 2c,d.
\par \smallskip
Table \ref{tab4} presents the light Higgs mass in the
approximation of Eq. (\ref{leadom2}) for the parameters of Figs. 2c,d.
\begin{table}[ht]
\begin{center}
\begin{tabular}{l||r|r||r|r}
 $\mu$ [GeV] & $M_{h,\, eff}$ [GeV] & $M_{h,\, 1-loop}$ [GeV]
 & $M_{h,\, eff}$ [GeV] &  $M_{h,\, 1-loop}$ [GeV]  \\
 \hline & & & & \\
& ($\tan \beta = 30$) & ($\tan \beta = 30$) &
 ($\tan \beta = 30$) & ($\tan \beta = 30$) \\
 & ($A= 0$) & ($A = 0$ GeV) & ($A = 300$) &
   ($A = 300$ GeV) \\
 -250  &  98.7   &   97.7  & 109.5   & 106.9   \\
 -100  & 100.4   &   98.1  & 111.2   & 108.4   \\
  0    & 100.4   &   98.1  & 111.5   & 108.8   \\
 100   & 100.4   &   98.0  & 111.5   & 109.0   \\
 250   &  98.7   &   96.5  & 110.3   & 107.4   \\
 \hline & & & & \\
& ($\tan \beta = 2$) & ($\tan \beta = 2$) &
 ($\tan \beta = 2$) & ($\tan \beta = 2$) \\
 & ($A= 0$) & ($A = 0$ GeV) & ($A = 300$) &
   ($A = 300$ GeV) \\
 -250  &  68.8   &   66.2  &  68.6   &  66.2   \\
 -100  &  65.8   &   63.5  &  74.2   &  71.8   \\
  0    &  65.2   &   61.8  &  79.1   &  76.1   \\
 100   &  65.8   &   61.8  &  84.8   &  81.3   \\
 250   &  68.8   &   64.4  &  92.2   &  89.2   \\
\end{tabular}
\end{center}
\caption{Physical light Higgs mass in the
approximation of Eq. (4.3) $M_{h,\, eff}$
and in the full one-loop result $M_{h,\, 1-loop}$.}
\label{tab4}
\end{table}
\par
The data from the approximation formulae are larger (smaller) by $< 1$-$4$ GeV
than the full result of Figs. 2c,d, depending in detail on the chosen
parameters.
\newpage
\par \smallskip
Slepton masses and slepton mixing angles show numerically small effects in
the neutral scalar Higgs mass. The variation of the slepton
soft breaking parameter $M_{\tilde{l}}$ between $100$ GeV
$< M_{\tilde{l}} < $ $500$ GeV while keeping squark masses
constant results in shifts of the light and
heavy Higgs mass by $0.3$ GeV. \\[0.3cm]
$\bullet$ Higgs mass dependence on gauginos
\medskip \par
Gauge bosons and gauginos in the virtual states of self energies and
tadpole diagrams are described in the MSSM by the SUSY soft breaking
parameters $\mu$, $M$, $M'$ and $\tan \beta$. In the numerical analysis
$M' = 5/3 \tan \theta_W \ M$.
In Figs. 3a,b the light Higgs mass is plotted for $m_t = 175$ GeV,
$M_A = 200$ GeV,
$\tan \beta = 2 (30)$ and $m_{sf}= 500$ GeV and a diagonal
sfermion mass matrix. The
values for $\mu$ and $M$ are in agreement with the experimentally lower
gaugino mass bounds.
The variation of the light Higgs mass dependence with $M$ in Figs. 3a,b is
below $2$ GeV and decreases for larger $M$ values.
By the choice of $\mu$, $A_t$ and $A_b$ are fixed in terms of Eq.
(\ref{sfbed}).
However, in Fig. 3b $A_b$ can reach its maximum value of $7.5$ TeV
and is substantially larger than all other SUSY soft breaking parameters.
Both plots show spikes as an effect of threshold effects of gauginos in
the various self energies. These effects can not be described by a
perturbation expansion and have to be skipped from the numerical
analysis.
\par
\bigskip
The full one-loop calculation of the neutral Higgs mass spectrum is
in good agreement with the simpler approximation formulae in most of
the parameter space. The deviations
between the full one-loop calculation and
the approximate result are within $2-10$ GeV in the considered
parameter regions.\\[0.3cm]
$\bullet$ Comparison with the complete on-shell renormalization scheme
\cite{pok1}
\par \smallskip
The full one-loop calculation of \cite{pok1} makes use of different
renormalization conditions.
\cite{pok1} fixes the counter terms $\delta v_i$ by an $\overline{MS}$
subtraction:
\begin{equation}
 \frac{\delta v}{v} = \frac{\delta v_i}{v_i} =
 - \frac{1}{( 4 \pi)^2} \frac{ (3 g_2^2 + g_1^2 ) }{4} \ \Delta ,
\label{dvpok}
\end{equation}
where the UV singularity $\Delta$ is given in (A.12).
Instead of (\ref{dvpok}) the renormalization condition for the residue
of the pseudoscalar Higgs propagator is used:
$$
 \Re e\ \frac{\partial}{\partial k^2} \hat{\Sigma}_{A^0} ( k^2 )  \mid _{
 k^2 = m_A^2} =  0 \ ,
$$
The numerical results of \cite{pok1} for the neutral Higgs masses, however,
are in agreement within
$0.4 \%$ with the one-loop calculation described in this article.\\[0.1cm]
\par
\bf Acknowledgements.\\ \rm
\smallskip
I am grateful to W. Hollik for various discussions and reading the
manuscript and to P.H. Chankowski, S. Pokorski, J. Rosiek for
high accuracy comparision and fruitful collaboration.
\appendix
\renewcommand{\thesubsection}{A}
\renewcommand{\theequation}{A.\arabic{equation}}
\subsection{Self energies and Tadpoles}
\setcounter{equation}{0}\setcounter{footnote}{0}
\vspace*{0.1cm} \hspace*{0.4cm}
The Feynman rules of the minimal supersymmetric standard model are given
in \cite{hunter}.
All analytical formulae are calculated in the 't Hooft-Feynman gauge.
The one- and two-point functions $A$, $B_0$, $B_1$, $B_{22}$ are defined at
the end of appendix A.
The self energies and tadpoles are split into the 2 Higgs doublet SM-,
sfermion- and gaugino contributions.
\smallskip \par
$\bullet \ Z^0$ - self energy:
\begin{eqnarray*}
 \lefteqn{
 \Sigma_{Z, 2-Higgs } (k^2)  =  \frac{\alpha}{4 \pi} \ \frac{1}{s_W^2 c_W^2}
 \ \left( \  - \sin^2 (\beta - \alpha ) ( \ B_{22} (k^2,M_A,m_{H^0})
 + B_{22} (k^2,M_Z,m_{h^0}) \ ) \ \right.  } \nonumber \\ & &
 - \cos^2 (\beta - \alpha ) ( \ B_{22} (k^2,M_A,m_{h^0})
 + B_{22} (k^2,M_Z,m_{H^0}) \ ) + 2 c_W^4 B_{22} (k^2,M_W,M_W) \nonumber \\
 & &  - \cos^2 2 \theta_W ( \ B_{22} (k^2,m_{H^+},m_{H^+}) + B_{22}
 (k^2,M_W,M_W) \ )  + \frac{1}{4} \ ( A (m_{h^0}) + A (m_{H^0})
 \nonumber \\ & &  + A (M_{A})
 + A (M_Z)  ) + \frac{ \cos^2 2 \theta_W }{2} ( A (M_W) +
 A (m_{H^+}) ) +  M_Z^2 ( 2 s_W^4 c_W^2 B_0 (k^2,M_W,M_W) \nonumber \\ & & +
 \sin^2 (\beta - \alpha ) B_0
 (k^2,M_Z,m_{h^0})  + \cos^2 (\beta - \alpha ) B_0 (k^2,M_Z,m_{H^0}) \ ) +
 c_W^4 ( A (M_W)  \nonumber \\ & &
 - 4 M_W^2 ) - c_W^4 F_2 (k^2, M_W, M_W)
 \nonumber \\ & &   -
 s_W^2 c_W^2 \sum_f \ N_C \ F_1 (k^2, m_f,m_f,\frac{v_f+a_f}{2},
 \frac{v_f-a_f}{2},\frac{v_f+a_f}{2},\frac{v_f-a_f}{2} \left. ) \ \right)
 \nonumber \\
%
 \lefteqn{
 \Sigma_{Z, sfermion} (k^2) =  - \frac{\alpha}{4 \pi} \
 \sum_{f} \  N_{C} \left( \ \frac{4  ( I_3^\sigma c_{\tilde{\theta}}^2
- Q_f s_W^2 )^2 }{s_W^2
 c_W^2} B_{22} (k^2,m_{\tilde{f}_1},m_{\tilde{f}_1})
 +  \frac{c_{\tilde{\theta}}^2 s_{\tilde{\theta}}^2 }{s_W^2
 c_W^2}  \right. }
 \nonumber \\ & &
 ( \ B_{22} (k^2,m_{\tilde{f}_1},m_{\tilde{f}_2}) +
 B_{22} (k^2,m_{\tilde{f}_2},m_{\tilde{f}_1}) \ )
 +  \frac{4  ( I_3^\sigma s_{\tilde{\theta}}^2
- Q_f s_W^2 )^2 }{s_W^2
 c_W^2} B_{22} (k^2,m_{\tilde{f}_2},m_{\tilde{f}_2}) \nonumber \\ & &
 \left. - 2 \frac{ ( I_3^\sigma - Q_f s_W^2)^2 c_{\tilde{\theta}}^2
 + 4 Q_f^2 s_W^4 s_{\tilde{\theta}}^2}{s_W^2 c_W^2 } A
 (m_{\tilde{f}_1})  - 2
 \frac{ ( I_3^\sigma - Q_f s_W^2)^2 s_{\tilde{\theta}}^2
 + 4 Q_f^2 s_W^4 c_{\tilde{\theta}}^2}{s_W^2 c_W^2 } A
 (m_{\tilde{f}_2}) \ \right)  \nonumber \\
 \lefteqn{
 \Sigma_{Z, gaugino} (k^2)  =  - \frac{\alpha}{4 \pi} \
 \frac{1}{4 s_W^2 c_W^2} \ \left( \ \frac{1}{2} \ \sum_{i,j = 1}^4 \ F_1 (
 k^2,m_{\tilde{\chi}_i^0},m_{\tilde{\chi}_j^0},O_{ij}''^L,
   O_{ij}''^R,O_{ji}''^L,O_{ji}''^R ) \right. } \nonumber \\ & & +
 \left. \sum_{i,j = 1}^2 F_1 (
 k^2,m_{\tilde{\chi}_i^+},m_{\tilde{\chi}_j^-},O_{ij}'^L,
   O_{ij}'^R,O_{ji}'^L,O_{ji}'^R ) \ \right) \ ,
\end{eqnarray*}
with the vector and axial vector coupling notation:
\begin{equation}
v_f = \frac{I_3^\sigma - 2 s_W^2 Q_f}{2 c_W s_W} \ , \
a_f = \frac{I_3^\sigma}{2 c_W s_W} \ .
\end{equation}
$I_3^\sigma = \pm 1/2$ is the weak isospin and $Q_f$ the fermion
charge and  $c_{\tilde{\theta}} = \cos \tilde{\theta}$,
$s_{\tilde{\theta}} = \sin \tilde{\theta}$.
The chargino and neutralino couplings are:
\begin{eqnarray}
 O_{ij}'^L = - V_{i1} V_{j1}^* - \frac{1}{2} V_{i2} V_{j2}^* +
                \delta_{ij} s_W^2 & ; &
 O_{ij}'^R  =  - U_{i1}^* U_{j1} - \frac{1}{2} U_{i2}^* U_{j2} +
                \delta_{ij} s_W^2 \nonumber \\
 O_{ij}''^L = - \frac{1}{2} N_{i3} N_{j3}^* + \frac{1}{2}
                N_{i4} N_{j4}^* & ; &
 O_{ij}''^R  =  - O_{ij}''^{L*} \ \ ,
\end{eqnarray}
where the chargino matrix $U_{ij}$, $V_{ij}$ and neutralino matrix
$N_{ij}$ are given in appendix B.
\smallskip \par
$\bullet \ W^\pm$ - self energy:
\begin{eqnarray*}  \lefteqn{
 \Sigma_{W, 2-Higgs } (k^2)  =  \frac{\alpha}{4 \pi} \ \frac{1}{s_W^2} \
 \left( \ - \sin^2 (\beta - \alpha ) ( \ B_{22} (k^2,m_{H^+},m_{H^0})
 + B_{22} (k^2,M_W,m_{h^0}) \ ) \right. } \\ & &
 - \cos^2 (\beta - \alpha ) ( \ B_{22}
 (k^2,m_{H^+},m_{h^0}) + B_{22} (k^2,M_W,m_{H^0}) \ ) + 2 s_W^2
 B_{22} (k^2, 0, M_W) \nonumber
 \\ & &  - \ B_{22} (k^2,M_{W},M_{Z}) - B_{22} (k^2,m_{H^+},M_A) + 2 c_W^2
 B_{22} (k^2,M_{W},M_{Z}) \nonumber \\ & &
  + \frac{1}{4} \ ( \ A (m_{h^0}) + A (m_{H^0}) + A (M_{A})
  + A (M_Z) + 2 A (M_W) + 2  A (m_{H^+})\ ) + M_W^2 \nonumber \\ & &
 ( \sin^2 (\beta - \alpha ) B_0 (k^2,M_W,m_{h^0})
  + \cos^2 (\beta - \alpha ) B_0 (k^2,M_W,m_{H^0}) + s_W^2
  B_0 (k^2,0,M_W) \nonumber \\ & & + \frac{s_W^4}{c_W^2} B_0 (k^2,M_Z,M_W) )
 + 3 A (M_W^2)  - 2 M_W^2 + c_W^2 ( 3 A (M_Z^2)  - 2 M_Z^2 ) \nonumber \\ &
 & - c_W^2 F_2 (k^2, M_Z, M_W) - s_W^2  F_2 (k^2, 0, M_W)
 - \sum_{f} \frac{N_C}{2} F_1 (k^2,m_{f^+},m_{f^-},
 \frac{1}{2},0,\frac{1}{2},0 \left. ) \  \right) \nonumber \\
\lefteqn{
 \Sigma_{W, sfermion} (k^2)  =   - \frac{\alpha}{4 \pi} \ \frac{1}{s_W^2}
 \ \sum_{f} \ N_{C} \ \left( \ 2 c_{\tilde{\theta}^+}^2
 c_{\tilde{\theta}^-}^2 B_{22} (k^2,m_{\tilde{f}_1^+},m_{\tilde{f}_1^-})
 + 2 c_{\tilde{\theta}^+}^2 s_{\tilde{\theta}^-}^2
 \right. } \nonumber \\ & &
 B_{22} (k^2,m_{\tilde{f}_1^+},m_{\tilde{f}_2^-})
 + 2 s_{\tilde{\theta}^+}^2 c_{\tilde{\theta}^-}^2
 B_{22} (k^2,m_{\tilde{f}_2^+},m_{\tilde{f}_1^-})
+ 2 s_{\tilde{\theta}^+}^2 s_{\tilde{\theta}^-}^2
 B_{22} (k^2,m_{\tilde{f}_2^+},m_{\tilde{f}_2^-})
 \nonumber \\ & &
 - \frac{
 c_{\tilde{\theta}^+}^2}{2} \ (A (m_{\tilde{f}_1^+}) + A (m_{\tilde{f}_1^-}) )
 - \frac{ s_{\tilde{\theta}^+}^2  }{2} \ ( A (m_{\tilde{f}_2^+})
 + A (m_{\tilde{f}_2^-})\left. ) \  \right)
  \nonumber \\
 \lefteqn{
 \Sigma_{W, gaugino} (k^2)  =  - \frac{\alpha}{4 \pi} \
 \frac{1}{4 s_W^2 } \sum_{i = 1}^{2} \sum_{j = 1}^{4}  \ F_1 (
 k^2,m_{\tilde{\chi}_i^+},m_{\tilde{\chi}_j^0},O_{ij}^L,
   O_{ij}^R,O_{ij}^L,O_{ij}^R ) \ , }
\end{eqnarray*}
with
\begin{eqnarray}
 O_{ij}^L =  - \frac{1}{\sqrt{2}} N_{j4} V_{i2}^* + N_{j2}
 V_{i1}^*   & ; &
 O_{ij}^R = + \frac{1}{\sqrt{2}} N_{j3}^* U_{i2} +  N_{j2}^* U_{i1}  \ \ .
\nonumber \\
\end{eqnarray}
\par
$\bullet \ $ Photon - self energy:
\begin{eqnarray*}
\lefteqn{
 \Sigma_{\gamma, 2-Higgs } (k^2)  =  - \frac{\alpha}{4 \pi} \ \left( \
 4  B_{22} (k^2,m_{H^+},m_{H^-}) + 2 B_{22} (k^2,M_{W},M_{W})
 \right. }  \nonumber \\ & &
 - 2 M_W^2 B_0 (k^2,M_{W},M_{W})
 - 2 A (M_{H^+}) - 8 A (M_W) + 4 M_W^2
 + F_2 (k^2,M_{W},M_{W}) \nonumber \\ & &
 + \sum_{f} \ N_C Q_f^2 \ F_1 (k^2, m_f, m_f, 1/2, 1/2, 1/2, 1/2
 \left. ) \ \right)
 \nonumber \\
\lefteqn{
 \Sigma_{\gamma, sfermion} (k^2) = - \frac{\alpha}{4 \pi} \ 2 \
 \sum_{f} \ N_{C} Q_f^2 \ ( \ 2
 B_{22} (k^2,m_{\tilde{f}_1},m_{\tilde{f}_1}) -  A
 (m_{\tilde{f}_1})  } \nonumber  \\ & & +
 \ 2 B_{22} (k^2,m_{\tilde{f}_2},m_{\tilde{f}_2}) -  A
 (m_{\tilde{f}_2}) \ )
 \nonumber \\
\lefteqn{
 \Sigma_{\gamma, gaugino} (k^2)  =  - \frac{\alpha}{4 \pi} \ \sum_{i =
 1}^{2}  F_1 ( k^2,m_{\tilde{\chi}_i^+},m_{\tilde{\chi}_i^-},1/2,1/2,1/2,1/2
) } \nonumber \\
\end{eqnarray*}
\par
$\bullet \ \gamma Z^0$ - mixing:
\begin{eqnarray*} \lefteqn{
 \Sigma_{\gamma Z, 2-Higgs } (k^2) = - \frac{\alpha}{4 \pi} \ \left( \
 \frac{2 \cos 2 \theta_W}{s_W c_W} ( \ B_{22} (k^2,m_{H^+},m_{H^-}) +
 B_{22} (k^2,M_W,M_W) \ ) \right.  }\nonumber \\ & &
 + \frac{2 c_W}{s_W} B_{22} (k^2,M_W,M_W) - 2 s_W c_W  M_Z^2
 B_0 (k^2,M_W,M_W) \nonumber \\ & &
 - \frac{ \cos 2 \theta_W}{s_W c_W} ( \ A (m_{H^+}) + A (M_W) \ )
 + \frac{c_W}{s_W} ( 6 A (M_W) - 4 M_W^2 ) - \frac{c_W}{s_W}  \nonumber
\\ &  & \left. F_2 (k^2,M_W,M_W) - \sum_f \ N_C Q_f \
 F_1 (k^2,m_f,m_f,\frac{1}{2},\frac{1}{2},\frac{v_f - a_f}{2},
 \frac{v_f + a_f}{2} ) \right)
 \nonumber \\
\lefteqn{
 \Sigma_{\gamma Z, sfermion} (k^2)  =   - \frac{\alpha}{4 \pi} \ 2 \
 \sum_{f} \ N_{C} Q_f \ \left( \frac{I_3^\sigma c_{\tilde{\theta}}^2
 - Q_{f} s_W^2 }{s_W c_W} \ ( 2
 B_{22} (k^2,m_{\tilde{f}_1},m_{\tilde{f}_1}) \right. } \nonumber \\ & &
 \left. - A  (m_{\tilde{f}_1}) \ ) +
 \frac{ I_3^\sigma s_{\tilde{\theta}}^2 - Q_{f} s_W^2 }{s_W c_W}
 \ ( 2 B_{22} (k^2,m_{\tilde{f}_2},m_{\tilde{f}_2}) -
 A (m_{\tilde{f}_2})\ ) \ \right)
 \nonumber \\
\lefteqn{
 \Sigma_{\gamma Z, gaugino} (k^2)  =  - \frac{\alpha}{4 \pi} \
 \frac{1}{2 s_W c_W} \ \sum_{i = 1}^{2} \ F_1 (
 k^2,m_{\tilde{\chi}_i^+},m_{\tilde{\chi}_i^-},1/2,1/2,O_{ii}'^L,O_{ii}'^R )
 } \nonumber \\
\end{eqnarray*}
The fermion loop contributions of vector boson self energies are
described by the following functions with masses and couplings
in its arguments.
\begin{eqnarray}
  F_1(k^2,m_1,m_2,a,b,a',b') & = & 8 \ [ \ ( a a' + b b' ) ( - 2 B_{22} +
  A (m_2) + m_1^2 B_0 + k^2 B_1 ) \nonumber \\ & & - \ ( a b' + a' b ) \
  m_1   m_2 \ B_0 \ ] \  (k^2,m_1,m_2) \nonumber \\
  F_2(k^2,m_1,m_2)  & = & [ \ 10 B_{22} + ( 4 k^2 + m_1^2 + m_2^2 ) B_0
  + A (m_1) + A (m_2) \nonumber \\ & & - 2 ( m_1^2 + m_2^2 - \frac{k^2}{3} )
  \ ]   \ (k^2,m_1,m_2) \ .
\end{eqnarray}
\smallskip \par
$\bullet \ $ Pseudoscalar Higgs self energy:
\begin{eqnarray*} \lefteqn{
\Sigma_{A,2-Higgs} ( k^2 ) = \frac{\alpha}{4 \pi} \ \frac{1}{ s_W^2} \
\left( \
 \frac{1}{2} S_1 (k^2,m_{H^+},M_W) + \frac{ \sin^2 ( \beta -
 \alpha) }{4 c_W^2 } S_1 (k^2,m_{H^0},M_Z)  \right. } \nonumber \\ & &
 + \frac{ \cos^2 ( \beta - \alpha) }{4 c_W^2} S_1 (k^2,m_{h^0},M_Z)
 + \frac{M_Z^2 \cos^2 2 \beta}{4 c_W^2} ( \cos^2 ( \beta + \alpha )
 B_0 (k^2,M_A,m_{H^0}) \nonumber \\ & & + \sin^2 ( \beta + \alpha ) B_0
 (k^2,M_A,m_{h^0}) ) + \frac{M_Z^2 \sin^2 2 \beta}{4 c_W^2} (
 \cos^2 ( \beta + \alpha ) B_0 (k^2,M_Z,m_{H^0}) \nonumber \\ & & +
 \sin^2 ( \beta + \alpha ) B_0 (k^2,M_Z,m_{h^0}) ) +
 \frac{M_W^2}{2} B_0 (k^2,M_W,m_{H^+}) \nonumber \\ & & +
 (  2 A (M_W) - M_W^2 ) + \frac{1}{2 c_W^2} ( 2 A (M_Z) -
 M_Z^2 ) + \frac{ \cos^2 2 \beta}{4 c_W^2} A (m_{H^+}) \nonumber \\
 & & + \frac{ 1 + \sin^2 2 \beta - \tan^2 \theta_W \cos^2 2 \beta }{4}
 A (M_{W})  + \frac{3 \cos^2 2 \beta}{8 c_W^2} A (M_{A}) \nonumber
 \\ & & + \frac{3 \sin^2 2 \beta - 1}{8 c_W^2} A (M_{Z}) +
 \frac{ \cos 2 \beta \cos 2 \alpha}{8 c_W^2} ( A (m_{h^0}) -
 A (m_{H^0}) ) \   \nonumber \\ &  & + \
 \sum_{f}  N_{C}  ( \ \frac{m_{f^-}^2 \tan^2 \beta}
 {4  M_W^2 } S_2 (k^2,m_{f^-},m_{f^-},-1/2,1/2,-1/2,1/2) \nonumber \\ & &
 \left. + \frac{m_{f^+}^2 \cot^2 \beta}{4 M_W^2 }
 S_2 (k^2,m_{f^+},m_{f^+},-1/2,1/2,-1/2,1/2) \ ) \ \right) \nonumber \\
\lefteqn{
 \Sigma_{A, sfermion} (k^2) = - \frac{\alpha}{4 \pi} \
 \sum_{f} \ N_{C} \ \left( \  -\frac{ m_{f^-}^2 (
 \mu - A_d \tan  \beta )^2 }{2 s_W^2 M_W^2 } B_0
 (k^2,m_{\tilde{f}_1^-},m_{\tilde{f}_2^-}) \right. } \nonumber \\ \hfill{
 } & & + (  \frac{ I_3^\sigma
 - Q_{-} s_W^2}{ 2 s_W^2 c_W^2} \cos 2 \beta - \frac{ m^2_{-} }
 {2 s_W^2 M_W^2} \tan^2 \beta ) \ ( c_{\tilde{\theta}}^2
 A (m_{\tilde{f}_1^-}) + s_{\tilde{\theta}}^2 A (m_{\tilde{f}_2^-}) \ )
 \nonumber \\ &
 & + ( \frac{
 Q_{-}  \tan^2 \theta_W}{2 s_W^2} \cos 2 \beta - \frac{ m^2_{-}
 }  {2 s_W^2 M_W^2} \tan^2 \beta ) \ ( c_{\tilde{\theta}}^2
 A (m_{\tilde{f}_2^-}) + s_{\tilde{\theta}}^2 A (m_{\tilde{f}_1^-}) \ )
 \nonumber
 \\ & &  - \frac{ m_{f^+}^2 ( \mu - A_u \cot \beta )^2 }{2 s_W^2 M_W^2 }
 B_0 (k^2,m_{\tilde{f}_1^+},m_{\tilde{f}_2^+}) \nonumber \\
 & & +
 ( \frac{ I_3^\sigma - Q_{+} s_W^2}{ 2 s_W^2 c_W^2} \cos 2 \beta
 - \frac{ m^2_{f^+} }{2 s_W^2 M_W^2} \cot^2 \beta ) \ ( c_{\tilde{\theta}}^2
 A (m_{\tilde{f}_1^+}) + s_{\tilde{\theta}}^2 A (m_{\tilde{f}_2^+}) \ )
 \nonumber \\ & & + \left.
 ( \frac{ Q_{+}  \tan^2 \theta_W}{2 s_W^2} \cos 2 \beta
 -\frac{ m^2_{f^+}}  {2 s_W^2 M_W^2} \cot^2 \beta ) \ ( c_{\tilde{\theta}}^2
 A (m_{\tilde{f}_2^+}) + s_{\tilde{\theta}}^2 A (m_{\tilde{f}_1^+}) \ )
 \ \right) \nonumber \\
\lefteqn{
 \Sigma_{A, gaugino} (k^2) = \frac{\alpha}{4 \pi} \frac{1}{4 s_W^2} \ \left(
 \ \sum_{i,j=1}^2 \ S_2
 (k^2,m_{\tilde{\chi}_i^+},m_{\tilde{\chi}_j^-},
 O_{ij},-O_{ji}^{*},O_{ji},-O_{ij}^{*})
  + \right. } \nonumber \\ & &
 \frac{1}{ 2 } \left. \ \sum_{i,j=1}^4 \ S_2
(k^2,m_{\tilde{\chi}_i^0},m_{\tilde{\chi}_j^0},O'_{ij},-O_{ji}^{'*},O'_{ji},-O_{ij}^{'*}
 )  \ \right) \ .
\end{eqnarray*} \\
The Higgs self energy contains the following functions:
\begin{eqnarray}
 S_1 (k^2,m_1,m_2) & = & - \ [ \ ( k^2 + m_1^2 ) B_0 - 2 k^2 B_1 +
   A (m_2) \ ] \ ( k^2, m_1,m_2) \nonumber \\
 S_2 (k^2,m_1,m_2,a,b,a',b') & = & 8 \ [ \ ( a b' + b a' ) (k^2 B_1 +
 A (m_1) + m_2^2 B_0 ) + \nonumber \\ & & \ ( a a' + b b' ) m_1 m_2 B_0 \ ]
 \  ( k^2, m_1,m_2) \nonumber \\
 S_3 (k^2,m) & = & [ \ 2 m^2 B_0 + A (m) + k^2 B_1 \ ] \ ( k^2,m,m ) \ ,
 \nonumber
\end{eqnarray}
together with the mixing coefficients of charginos and neutralinos
to the pseudoscalar Higgs
\begin{eqnarray}
 O_{ij}  & = & Q_{ij} \sin \beta + S_{ij} \cos \beta  \nonumber \\
 O_{ij}' & = & Q_{ij}'' \sin \beta - S_{ij}'' \cos \beta  \nonumber
\end{eqnarray}
\begin{equation}
 Q_{ij} = \sqrt{\frac{1}{2}} \ V_{i1} \ U_{j2} \ , \
 S_{ij} = \sqrt{\frac{1}{2}} \ V_{i2} \ U_{j1}   \\
\end{equation}
\begin{eqnarray}
 Q_{ij}'' & = & \frac{1}{2} \ [ \ N_{i3} ( N_{j2} - N_{j1} \tan \theta_W ) +
        N_{j3} ( N_{i2} - N_{i1} \tan \theta_W ) \ ] \nonumber \\
 S_{ij}'' & = & \frac{1}{2} \ [ \ N_{i4} ( N_{j2} - N_{j1} \tan \theta_W ) +
        N_{j4} ( N_{i2} - N_{i1} \tan \theta_W ) \ ]  \nonumber \ .
\end{eqnarray} \\
\par
$\bullet \ \Sigma_{h^0}$, $\Sigma_{H^0}$ self energy and
$h^0H^0$ - mixing $\Sigma_{hH}$:
\begin{eqnarray*} \lefteqn{
 \Sigma_{\{h^0,H^0,hH\}, 2-Higgs} = \frac{\alpha}{4 \pi} \ \frac{1}{s_W^2}
 \left( \
 \frac{ \{ \cos^2,\sin^2,-\sin \cos \}  ( \beta - \alpha ) }{2} \ (
 S_1(k^2,m_{H^+},M_W) \right. } \nonumber \\ & & + \frac{1}{2 c_W^2}
S_1(k^2,M_A,M_Z)
)  + \frac{ \{ \sin^2,\cos^2,\sin \cos \}  ( \beta - \alpha ) }{2} \
 ( S_1(k^2,M_W,M_W) \nonumber \\ & &  + \frac{1}{2 c_W^2}
S_1(k^2,M_Z,M_Z)
)  +  M_W^2 \{ \sin^2,\cos^2,\sin \cos \} ( \beta - \alpha )
( 4 B_0 (k^2,M_W,M_W) - 2  \nonumber \\ & &  + \frac{1}{2
 c_W^4} \ ( 4 B_0  (k^2,M_Z,M_Z) - 2 ) )
 + M_W \{ c_{1,1}^2,c_{1,2}^2,c_{1,1} c_{1,2} \} \ B_0
 (k^2,m_{H^+},m_{H^+})
 \nonumber \\ & & + \frac{ \{s_{1,1}^2,s_{1,2}^2,s_{1,1}
 s_{1,2} \}}{2} \ B_0  (k^2,m_{h^0},m_{h^0}) +
 \{s_{1,2}^2,s_{2,2}^2,s_{1,2} s_{2,2}\} \ B_0
               (k^2,m_{h^0},m_{H^0})
 \nonumber \\ & & + \frac{\{s_{2,2}^2,s_{2,1}^2,s_{2,2}
 s_{2,1}\}}{2}
 \  B_0 (k^2,m_{H^0},m_{H^0}) + \frac{\{p_{1,1}^2,p_{1,2}^2,p_{1,1}
 p_{1,2}\}}{2} \
 ( B_0 (k^2,M_{A},M_{A}) \nonumber \\ & & + B_0
 (k^2,M_{Z},M_{Z})
 +  2 B_0 (k^2,M_{W},M_{W}) ) +  \{p_{2,1}^2,p_{2,2}^2,p_{2,1}
p_{2,2}\} B_0 (k^2,M_A,M_Z) \nonumber \\ & &
 + \frac{M_W^2 \{c_{2,1}^2,  c_{2,2}^2,c_{2,1} c_{2,2}\}}{2} \ B_0
 (k^2,M_W,m_{H^+})
-  \frac{M_W^2 \{ \sin^2, \cos^2, \sin \cos \} (\beta - \alpha ) }{2}
\nonumber \\ & & \cdot ( B_0 (k^2, M_W,M_W) + \frac{1}{2 c_W^4}
 B_0 (k^2,M_Z,M_Z) ) + \
 \{1,1,0\}\ ( \ 2 A(M_W) - M_W^2
 \nonumber \\ & & + \frac{1}{2 c_W^2}
 ( 2 A (M_Z) -  M_Z^2 ) \ )  + \frac{\{3 \cos^2 2 \alpha, 3 \sin^2 2 \alpha
 - 1, 3 \sin 2 \alpha \cos 2 \alpha \} }{8 c_W^2 } A (m_{h^0})
 \nonumber \\  & &
 + \frac{\{3 \sin^2 2 \alpha - 1,3 \cos^2 2 \alpha,- 3 \sin 2 \alpha
 \cos 2 \alpha \} }{8 c_W^2 } \ A (m_{H^0}) \nonumber \\ & &
 + \frac{\{\cos 2 \beta \cos 2 \alpha , - \cos 2 \beta \cos 2 \alpha ,
 \cos 2 \beta \sin 2 \alpha \}}{8 c_W^2 } \ ( \ A (M_A) - A (M_Z) \ )
 \nonumber \\ & & + \frac{
\{1+v_{1},1-v_{1},-v_{2} \}}{4}
 \ A (M_W) + \frac{\{1-v_{1},1+v_{1},v_{2}\}}{4} \ A
 (m_{H^+})  \nonumber \\
  & & -
 \sum_{f}  N_{C} \ ( \ \frac{m_{f^+}^2}{
  M_W^2 \sin^2 \beta } \{ \cos^2 \alpha , \sin^2 \alpha , \sin \alpha
 \cos \alpha \} \ S_3 (k^2,m_{f^+})  \nonumber \\
 & & \left. + \frac{m_{f^-}^2}{ M_W^2 \cos^2 \beta } \{ \sin^2 \alpha ,
 \cos^2 \alpha , - \sin \alpha \cos \alpha \} \ S_3 (k^2,m_{f^-}) \ ) \ \right)
 \nonumber \\
\lefteqn{
 \Sigma_{\{h^0,H^0,hH\}, sfermion} = \frac{\alpha}{4 \pi}
\frac{1}{s_W^2}
 \sum_{f} N_{C} \ \left( \ ( \ c_{\tilde{\theta}}^2 \{ u_{1,1},
 u_{2,1}, u_{1,1}  \} \right. } \nonumber \\ & & + s_{\tilde{\theta}}^2 \
  \{ u_{1,2}, u_{2,2}, u_{1,2}  \} +
 \sin 2 \tilde{\theta} \{ u_{1,3}, u_{2,3}, u_{1,3} \} \ )
\ ( \ c_{\tilde{\theta}}^2 \{ u_{1,1},
 u_{2,1}, u_{2,1} \}  \nonumber \\ & & + s_{\tilde{\theta}}^2 \
  \{ u_{1,2}, u_{2,2}, u_{2,2} \} +
 \sin 2 \tilde{\theta} \{ u_{1,3}, u_{2,3}, u_{2,3} \} )
 \
 \ B_0 (k^2, m_{\tilde{f}_1},m_{\tilde{f}_1} )
 \nonumber \\ & & + 2 \ ( \
 c_{\tilde{\theta}} s_{\tilde{\theta}} (
 \{ u_{1,2}, u_{2,2}, u_{1,2} \} -
 \{ u_{1,1}, u_{2,1}, u_{1,1} \} ) + \cos 2 \tilde{\theta}
 \nonumber \\ & &
 \{ u_{1,3}, u_{2,3}, u_{1,3} \} \ ) \ ( \
 c_{\tilde{\theta}} s_{\tilde{\theta}} (
 \{ u_{1,2}, u_{2,2}, u_{2,2} \} -
 \{ u_{1,1}, u_{2,1}, u_{2,1} \} ) + \cos 2 \tilde{\theta}
 \nonumber \\ & &
 \{ u_{1,3}, u_{2,3}, u_{2,3} \} \ )
 \ B_0 (k^2, m_{\tilde{f}_1},m_{\tilde{f}_2} ) +
\ ( \ s_{\tilde{\theta}}^2 \{ u_{1,1},
 u_{2,1}, u_{1,1} \}  \nonumber \\ & & + c_{\tilde{\theta}}^2 \
  \{ u_{1,2}, u_{2,2}, u_{1,2} \} -
 \sin 2 \tilde{\theta} \{ u_{1,3}, u_{2,3}, u_{1,3} \}
 \ ) \ ( \ s_{\tilde{\theta}}^2 \{ u_{1,1},
 u_{2,1}, u_{2,1} \}  \nonumber \\ & & + c_{\tilde{\theta}}^2 \
  \{ u_{1,2}, u_{2,2}, u_{2,2} \} -
 \sin 2 \tilde{\theta} \{ u_{1,3}, u_{2,3}, u_{2,3} \}
 \ ) \ B_0 (k^2, m_{\tilde{f}_2},m_{\tilde{f}_2} )
 \nonumber \\ & & - c_{\tilde{\theta}}^2
 ( \frac{ 1/2 - Q_+ s_W^2}{2 c_W^2} \{ \cos , -\cos  , \sin
 \}(2 \alpha)  - \frac{
 m_{f^+}^2  \{ \cos^2 , \sin^2 , \sin \cos \} ( \alpha ) }{2
 M_W^2 \sin^2 \beta } )  A ( m_{\tilde{f}_1^+} ) \nonumber \\ & & -
 s_{\tilde{\theta}}^2  ( \frac{-1/2 - Q_- s_W^2 }{2 c_W^2} \{ \cos , -\cos  ,
\sin  \}(2 \alpha ) - \frac
 { m_{f^-}^2  \{ \sin^2  , \cos^2 , - \sin \cos \} ( \alpha )
 }{2 M_W^2 \cos^2 \beta  }  ) A ( m_{\tilde{f}_1^-} ) \nonumber \\ & & -
 s_{\tilde{\theta}}^2 ( \frac{Q_+ s_W^2}{c_W^2} \{ \cos, - \cos , \sin
  \}(2 \alpha )  - \frac{
 m_{f^+}^2  \{ \cos^2 , \sin^2 ,\sin \cos \} ( \alpha ) }{2
 M_W^2 \sin^2 \beta} ) A ( m_{\tilde{f}_2^+}) \nonumber \\ & & - \left.
 c_{\tilde{\theta}}^2  ( \frac{Q_- s_W^2}{c_W^2} \{ \cos , - \cos ,
\sin  \} (2 \alpha ) - \frac{
 m_{f^-}^2  \{ \sin^2 , \cos^2 ,- \sin \cos \}( \alpha ) }{2
 M_W^2 \cos^2 \beta } )  A ( m_{\tilde{f}_2^-} ) \ \right) \nonumber \\
\lefteqn{
 \Sigma_{\{h^0,H^0,hH\}, gaugino} =  - \frac{\alpha}{4 \pi}
 \frac{1}{4 s_W^2} \ \left( \ \sum_{i,j=1}^{2} S_2 (
k^2,m_{\tilde{\chi}_i^+},m_{\tilde{\chi}_j^-},\{O_{ij}^h,O_{ij}^H,O_{ij}^h\},
 \right.  } \nonumber \\ & &
 \{O_{ji}^{h*},O_{ji}^{H*},O_{ji}^{h*}\} ,
\{O_{ji}^h,O_{ji}^H,O_{ji}^H\}, \{O_{ij}^{h*},O_{ij}^{H*},O_{ij}^{H*}\} )
 + \frac{1}{2} \sum_{i,j=1}^{4} S_2 (
 k^2,m_{\tilde{\chi}_i^0},m_{\tilde{\chi}_j^0}, \nonumber \\ & &
 \left.  \{O_{ij}^{'h},O_{ij}^{'H},O_{ij}^{'h}\},
 \{O_{ji}^{'h*},O_{ji}^{'H*},O_{ji}^{'h*}\},
 \{O_{ji}^{'h},O_{ji}^{'H},O_{ji}^{'H}\},
 \{O_{ij}^{'h*},O_{ij}^{'H*},O_{ij}^{'H*}\} ) \ \right) \nonumber \\
\end{eqnarray*}
with the mixing coefficients
\begin{eqnarray}
 O_{ij}^h & = &  Q_{ij} \sin \alpha - S_{ij} \cos \alpha \nonumber \\
 O_{ij}^H & = &  Q_{ij} \cos \alpha + S_{ij} \sin \alpha \nonumber \\
 O_{ij}^{'h} & = &  Q_{ij}'' \sin \alpha + S_{ij}'' \cos \alpha \nonumber \\
 O_{ij}^{'H} & = &  Q_{ij}'' \cos \alpha - S_{ij}'' \sin \alpha \ ,
\end{eqnarray}
and the notation for Higgs couplings
\begin{eqnarray}
 (c_{i,j}) & = & \left( \begin{array}{ll}
 \sin ( \beta - \alpha ) + \frac{ 1 }{2 c_W^2 }
                 \cos 2 \beta \sin ( \alpha + \beta ), &
 \cos ( \beta - \alpha ) - \frac{ 1 }{2 c_W^2 }
                 \cos 2 \beta \cos ( \alpha + \beta ) \\
 \cos ( \beta - \alpha ) - \frac{1}{c_W^2}
               \sin 2 \beta \sin ( \alpha + \beta ), &
 - \sin ( \beta - \alpha ) + \frac{1}{c_W^2}
               \sin 2 \beta \cos ( \alpha + \beta ) \\
 \end{array} \right) \nonumber \\[0.3cm]
 (s_{i,j}) & = & \left( \begin{array}{ll}
 \frac{3 M_Z}{2 c_W} \cos 2 \alpha \sin ( \alpha + \beta ), &
 \frac{M_Z}{2 c_W} ( 2 \sin 2 \alpha \sin ( \alpha +
    \beta ) - \cos ( \alpha + \beta ) \cos 2 \alpha )  \\
 \frac{3 M_Z}{2 c_W} \cos 2 \alpha \cos ( \alpha + \beta ), &
 -\frac{M_Z}{2 c_W} ( 2 \sin 2 \alpha \cos
    ( \alpha + \beta ) + \sin  ( \alpha + \beta ) \cos 2 \alpha ) \\
  \end{array} \right) \nonumber \\[0.3cm]
 (p_{i,j}) & = & \left( \begin{array}{ll}
  \frac{M_Z}{2 c_W} \cos 2 \beta \sin ( \alpha + \beta ), &
  -\frac{M_Z}{2 c_W} \cos 2 \beta \cos ( \alpha +
                 \beta ) \nonumber \\
  \frac{M_Z}{2 c_W} \sin 2 \beta \sin ( \alpha + \beta ), &
  -\frac{M_Z}{2 c_W} \sin 2 \beta \cos ( \alpha +
               \beta )  \\
 \end{array} \right) \nonumber \\[0.3cm]
 (v_i) & = & \left(
  \sin 2 \alpha \sin 2 \beta - \tan^2 \theta_W
               \cos 2 \beta \cos 2 \alpha , \
  \sin 2 \beta \cos 2 \alpha + \tan^2 \theta_W
               \cos 2 \beta \sin 2 \alpha \right)  \nonumber \\[0.3cm]
 (u_{1,j}) & = & \left( \begin{array}{l}
  \frac{M_Z}{c_W} ( \pm \frac{1}{2} - Q_\pm s_W^2 )
         \sin ( \alpha + \beta ) - \frac{ m_\pm^2 \{ \cos \alpha , -
 \sin \alpha \} }{ M_W \{ \sin \beta , \cos \beta \} }  \\
  \frac{M_Z}{c_W} Q_\pm s_W^2
               \sin ( \alpha + \beta ) - \frac{ m_\pm^2 \{ \cos \alpha , -
 \sin \alpha \} }{ M_W \{ \sin \beta , \cos \beta \} }  \\
 \frac{m_f}{2 M_W \{ \sin \beta, \cos \beta \} }
 \{ \mu \sin \alpha - A_u \cos \alpha, \
    \mu \cos \alpha - A_d \sin \alpha \}  \\
 \end{array} \right) \nonumber \\[0.3cm]
 (u_{2,j}) & = & \left( \begin{array}{l}
  - \frac{M_Z}{c_W} ( \pm \frac{1}{2} - Q_\pm s_W^2 )
        \cos( \alpha + \beta ) - \frac{ m_\pm^2 \{ \sin \alpha ,
        \cos \alpha \} }{ M_W \{ \sin \beta , \cos \beta \} } \\
  - \frac{M_Z}{c_W} Q_\pm s_W^2
         \cos( \alpha + \beta ) - \frac{ m_\pm^2 \{ \sin \alpha ,
         \cos \alpha \} }{ M_W \{ \sin \beta , \cos \beta \} }  \\
 \frac{- m_f}{2 M_W \{ \sin \beta, \cos \beta \} }
 \{ \mu \cos \alpha + A_u \sin \alpha , \
    \mu \sin \alpha + A_d \cos \alpha \}  \\
 \end{array} \right) \nonumber \\[0.3cm]
\end{eqnarray}
\par
$\bullet \ A^0$-$Z^0$ - mixing:
\begin{eqnarray*} \lefteqn{
 \Sigma_{AZ,gauge} = \frac{\alpha}{4 \pi} \ \frac{1}{2 s_W^2 c_W^2}
 \left( \ - M_Z^2 \cos
 ( \beta - \alpha ) \sin ( \beta - \alpha ) \ (
 M_1 (k^2,m_{h^0},M_Z) \right. }  \nonumber \\ & &
- M_1(k^2,m_{H^0},M_Z) )   -
 \frac{ M_Z^2 \cos 2 \beta }{2} \ ( \sin ( \beta + \alpha )
 \cos (\beta - \alpha)
 M_2 (k^2,M_A,m_{h^0}) \nonumber \\ & &  + \cos ( \beta +
 \alpha
 ) \sin   ( \beta - \alpha ) M_2 (k^2,M_A,m_{H^0}) \ ) \nonumber \\ & & +
 \frac{ M_Z^2 \sin 2 \beta }{2} \ ( \sin ( \beta + \alpha )
 \sin ( \beta - \alpha ) M_2 (k^2,M_Z,m_{h^0}) \nonumber \\ & & -
 \cos ( \beta + \alpha ) \cos ( \beta - \alpha ) M_2 (k^2,M_Z,m_{H^0}) \ )
  \nonumber \\ & & - \
 \sum_{f}  N_{C} \ ( \ \frac {m_{f^-} s_W c_W^2 \tan \beta }
 {M_W} M_3 (k^2,m_{f^-},m_{f^-},v_f + a_f
 ,v_f - a_f,-\frac{1}{2},\frac{1}{2} ) \nonumber \\ & & + \left.
 \frac {m_{f^+} s_W c_W^2 \cot \beta }{M_W} M_3
 (k^2,m_{f^+},m_{f^+},v_f + a_f,v_f - a_f
 ,- \frac{1}{2},\frac{1}{2} ) \ ) \ \right)
 \nonumber \\[0.2cm]
\lefteqn{
 \Sigma_{AZ,sfermion} =  0 } \nonumber \\
\lefteqn{
 \Sigma_{AZ,gaugino} = - \frac{\alpha}{4 \pi} \frac{1}{4 s_W^2 c_W} \ \left( \
 \sum_{i,j=1}^2 \ M_3 (k^2,m_{\tilde{\chi}_i^+},m_{\tilde{\chi}_j^-},
 O_{ij}',-O_{ji}^{'*},O_{ij}^{'L},O_{ji}^{'R} ) \right.  } \nonumber \\ & & +
 \left. \frac{1}{2}  \sum_{i,j=1}^{4} \ M_3
 (k^2,m_{\tilde{\chi}_i^0},m_{\tilde{\chi}_j^0},O_{ij}'',-O_{ji}'',
 O_{ij}^{''L},O_{ji}^{''R} ) \  \right) \nonumber \  . \\
\end{eqnarray*}
where the functions in the $A^0$-$Z^0$ mixing are:
\begin{eqnarray}
 M_1 (k^2,m_1,m_2) & = & - ( B_0 - B_1 ) (k^2,m_1,m_2)  \nonumber \\
 M_2 (k^2,m_1,m_2) & = & ( B_0 + 2 B_1 ) (k^2,m_1,m_2)  \nonumber \\
 M_3 (k^2,m_1,m_2,a,b,c,d) & = & - 8 (\ (a c + b d) \ m_2 \ ( B_0 + B_1 )
        \nonumber \\ & &  + \
    ( a d + b c ) \ m_1 \ B_1 \ ) (k^2,m_1,m_2) \nonumber \ , \\
\end{eqnarray}
$\bullet \ $ Higgs - tadpoles:
\begin{eqnarray*} \lefteqn{
 T_{\{h^0,H^0\},gauge} = \frac{e}{(4 \pi)^2} \frac{1}{s_W} \ \left( \
 - M_W \{ \sin , \cos \} (\beta - \alpha ) \ ( 4 A (M_W) -
 2 M_W^2  \right. } \nonumber \\
 & & + \frac{1}{2 c_W^2} \ ( 4 A (M_Z)
 - 2 M_Z^2 ) \ ) - M_W \{ c_{1,1} , c_{1,2} \} \ A (m_{H^+})
 - \frac{1}{2} \{ s_{1,1} , s_{1,2} \} \ A (m_{h^0}) \nonumber \\ & &
 - \frac{1}{2} \{ s_{2,2} , s_{2,1} \} A (m_{H^0})
 + \frac{1}{2} \{ p_{1,1} , p_{1,2} \} \ ( A (M_A) - A (M_Z) - 2
 A (M_W) )
\nonumber \\ & &  + M_W \{ \sin , \cos \} ( \beta - \alpha ) \, ( A (M_W) +
\frac{1}{2 c_W^2} A (M_Z) )  \nonumber \\ & &
 + \left. \sum_{f} N_{C} \ ( \ \frac{ 2 m_{f^+}^2 \{ \cos
 \alpha , \sin \alpha \} }{M_W \sin \beta} A (m_{f^+}) +
 \frac{ 2 m_{f^-}^2 \{ - \sin \alpha , \cos \alpha \} }{M_W \cos \beta}
 A (m_{f^-}) \ ) \ \right) \nonumber \\
\lefteqn{
 T_{\{h^0,H^0\},sfermion} = \frac{e}{(4 \pi)^2} \ \frac{1}{s_W} \
 \ \sum_{f} N_{C} \ \left( ( \ c_{\tilde{\theta}}^2
 \{ u_{1,1} , u_{2,1} \} + s_{\tilde{\theta}}^2
 \{ u_{1,2} , u_{2,2} \} + \sin 2 \tilde{\theta} \right. } \nonumber \\ & &
 \{ u_{1,3} , u_{2,3} \} \ ) A ( m_{\tilde{f}_1} ) +
 ( \ s_{\tilde{\theta}}^2 \{ u_{1,1} , u_{2,1} \} + c_{\tilde{\theta}}^2
 \{ u_{1,2} , u_{2,2} \} - \sin 2 \tilde{\theta}
 \{ u_{1,3} , u_{2,3} \} \ ) \nonumber \\ & & \left.
  A ( m_{\tilde{f}_2} ) \ \right)
  \nonumber \\
\lefteqn{
 T_{\{h^0,H^0\},gaugino} = -\frac{e}{4 \pi^2} \ \frac{1}{s_W} \ \left( \
 \sum_{i=1}^2 \ ( Q_{ii} \{ \sin \alpha , -\cos \alpha \} - S_{ii}
 \{ \cos \alpha ,  \sin \alpha \} ) \ m_{\tilde{\chi}_i^+}
 \right. } \nonumber \\ & & \left.
 A ( m_{\tilde{\chi}_i^+} )  + \frac{1}{2}
 \sum_{i=1}^4 \ ( Q_{ii}'' \{ \sin \alpha , - \cos \alpha \} + S_{ii}''
 \{ \cos \alpha , \sin \alpha \} ) \ m_{\tilde{\chi}_i^0} \
 A ( m_{\tilde{\chi}_i^0} ) \ \right) \nonumber  \ .  \\
\end{eqnarray*}
In the self energies and tapoles the scalar 1-point integral is given
by:
\begin{equation}
 A (m) = m^2 \left ( \Delta - \log \frac{m^2}{\bar{\mu}^2} + 1 \right ) \ .
\end{equation}
For the scalar 2-point integral we can write
 \begin{equation}
 B_0(q^2,m_1,m_2) = \Delta - \int_0^1 \, dx \,\log\frac
    {x^2q^2 -x(q^2+m_1^2-m_2^2)+m_1^2-i\epsilon}{\bar{\mu}^2}  \, .
 \end{equation}
Explicit analytic expression are presented in \cite{bhs}.
In terms of $B_0$, $A$ one obtains for $B_1$ and $B_{22}$:
\begin{eqnarray}
 B_1(q^2,m_1,m_2) & = & - \frac{q^2+m_1^2-m_2^2}{2q^2}\,B_0(q^2,m_1,m_2)
       +  \frac{m_1^2-m_2^2}{2q^2}\, B_0(0,m_1,m_2) \nonumber \\
 B_{22} (k^2,m_1,m_2) & = & \frac{1}{3} \ \left( \ \frac{1}{2} A (m_1)
 + m_1^2 B_0 (k^2,m_1,m_2) + \right. \nonumber \\
 & & \left. + \frac{1}{2} ( k^2 + m_1^2 - m_2^2 ) B_1 (k^2,m_1,m_2) +
 \frac{m_1^2 + m_2^2}{2} - \frac{k^2}{6} \ \right) \ .
\end{eqnarray}
The expression
\begin{equation}
\Delta = \frac{2}{\varepsilon} - \gamma + \log 4\pi, \;\;\;\;\;\;
    \varepsilon = 4-D \, ,
\label{deltadr}
\end{equation}
and the mass scale $\bar{\mu}$ are the conventional UV
parameters in dimensional regularization.
\renewcommand{\thesubsection}{B}
\renewcommand{\theequation}{B.\arabic{equation}}
\subsection{Gaugino mass matrix}
\setcounter{equation}{0}\setcounter{footnote}{0}
The chargino $2 \times 2$ mass matrix is given by
\begin{equation}
  \cal{M}_{\rm \tilde{\chi}^\pm} \rm = \left( \begin{array}{ll}
    M & M_W \sqrt{2} \sin \beta \\ M_W \sqrt{2} \cos \beta & \mu \\
    \end{array}  \right) \ ,
\end{equation} \\
with the SUSY soft breaking parameters $\mu$ and $M$ in the diagonal
matrix elements. The physical chargino mass states $\tilde{\chi}^{\pm}_i$
are the rotated
wino and charged Higgsino states:
\begin{eqnarray}
\tilde{\chi}^+_i & = & V_{ij} \psi^+_j   \nonumber \\
\tilde{\chi}^-_i & = & U_{ij} \psi^-_j  \ ; \ i,j = 1,2  \ .
\end{eqnarray}
$V_{ij}$ and $U_{ij}$ are unitary chargino mixing matrices obtained from
the diagonalization of the mass matrix (B.1):
\begin{equation}
\rm U^* \cal{M}_{\rm \tilde{\chi}^\pm} \rm  V^{-1} =
diag(m_{\tilde{\chi}^\pm_1}^2,m_{\tilde{\chi}^\pm_2}^2) \ .
\end{equation}
\smallskip \par
The neutralino $4 \times 4 $ mass matrix yields:
\begin{equation}
 \cal{M}_{\rm \tilde{\chi}^0}   = \left( \begin{array}{cccc}
 M' & 0 & - M_Z \sin \theta_W \cos \beta & M_Z \sin \theta_W \sin \beta \\
 0 & M & M_Z \cos \theta_W \cos \beta & - M_Z \cos \theta_W \sin \beta \\
- M_Z \sin \theta_W \cos \beta & M_Z \cos \theta_W \cos \beta & 0 & - \mu \\
 M_Z \sin \theta_W \sin \beta & - M_Z \cos \theta_W \sin \beta & - \mu & 0
 \\  \end{array} \right)
\end{equation}
where the diagonalization introduces the unitary matrix $N_{ij}$ by:
\begin{equation}
  \rm N^* \cal{M}_{\rm \tilde{\chi}^0} \rm N^{-1}  = diag(
  m_{\tilde{\chi}_i^0}) \ .
\end{equation}
%
%
\bigskip
\newpage
{\bf FIGURE CAPTIONS}
\vskip 0.5cm
\noindent {\bf Figure 1.}~
Complete one-loop result of
the light and heavy neutral Higgs masses $M_h$, $M_H$ for the input
parameters $M_A$ and  $\tan \beta = $ $0.5$, $2$, $5$, $10$, $30$
in Figs. 1a,b  with a top quark mass $m_t = 175$ GeV.
Figs. 1c,d. show the light and heavy Higgs masses for $150$ GeV $< m_t <$
$200$ GeV and various choices for $\tan \beta$ together with a fixed
pseudoscalar mass $M_A = 200$ GeV.
A sfermion soft breaking parameter
$m_{sf} = 500$ GeV and $\mu = 100$ GeV is chosen and absence of
left-right mixing is assumed in Figs. 1a,b,c,d. The gaugino
soft breaking parameter $M = 400$ GeV in all figures.
\vskip 0.5cm
\noindent {\bf Figure 2.}~
Dependence of the light and heavy Higgs masses $M_h$, $M_H$ on
sfermions for $100$ GeV $< m_{sf} <$ $1$ TeV and various $\tan \beta$
values in Figs. 2a,b. $\mu = 100$ GeV and no left-right mixing.
The effects of the left-right mixing is presented in Figs. 2c,d
for the light Higgs mass $M_h$ as a function of $\mu$ and $A = A_t = A_b =$
$0$, $100$, $200$, $300$ GeV. In Fig. 2c(d) $\tan \beta = 30$ ($2$).
$M_A = 200$ GeV, $m_t = 175$ GeV and $M = 400$ GeV.
\vskip 0.5cm
\noindent {\bf Figure 3.}~
Effects of the gaugino sector in the light Higgs mass $M_h$ in Figs. 3a(b)
for $\tan \beta = 2$ ($30$). $M_A = 200$ GeV, $m_t = 175$ GeV,
$m_{sf} = 500$ GeV, no left-right mixing.
\newpage
\renewcommand{\thepage}{}
\begin{figure}
\epsfig{figure=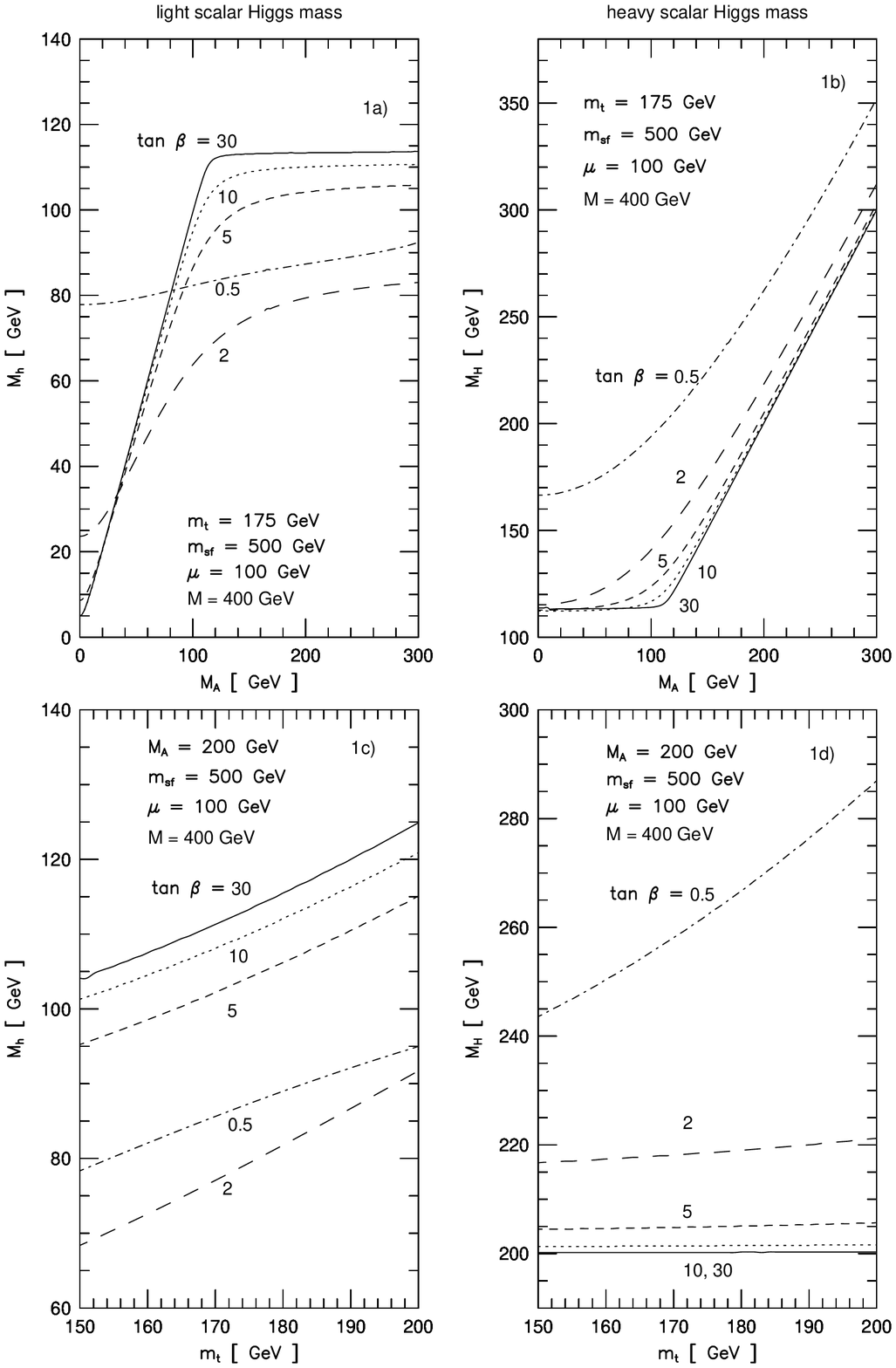,%
height=19cm,width=18.0cm,%
bbllx=75pt,bblly=75pt,bburx=555pt,bbury=751pt}
\caption{}
\end{figure}
\newpage
\begin{figure}
\epsfig{figure=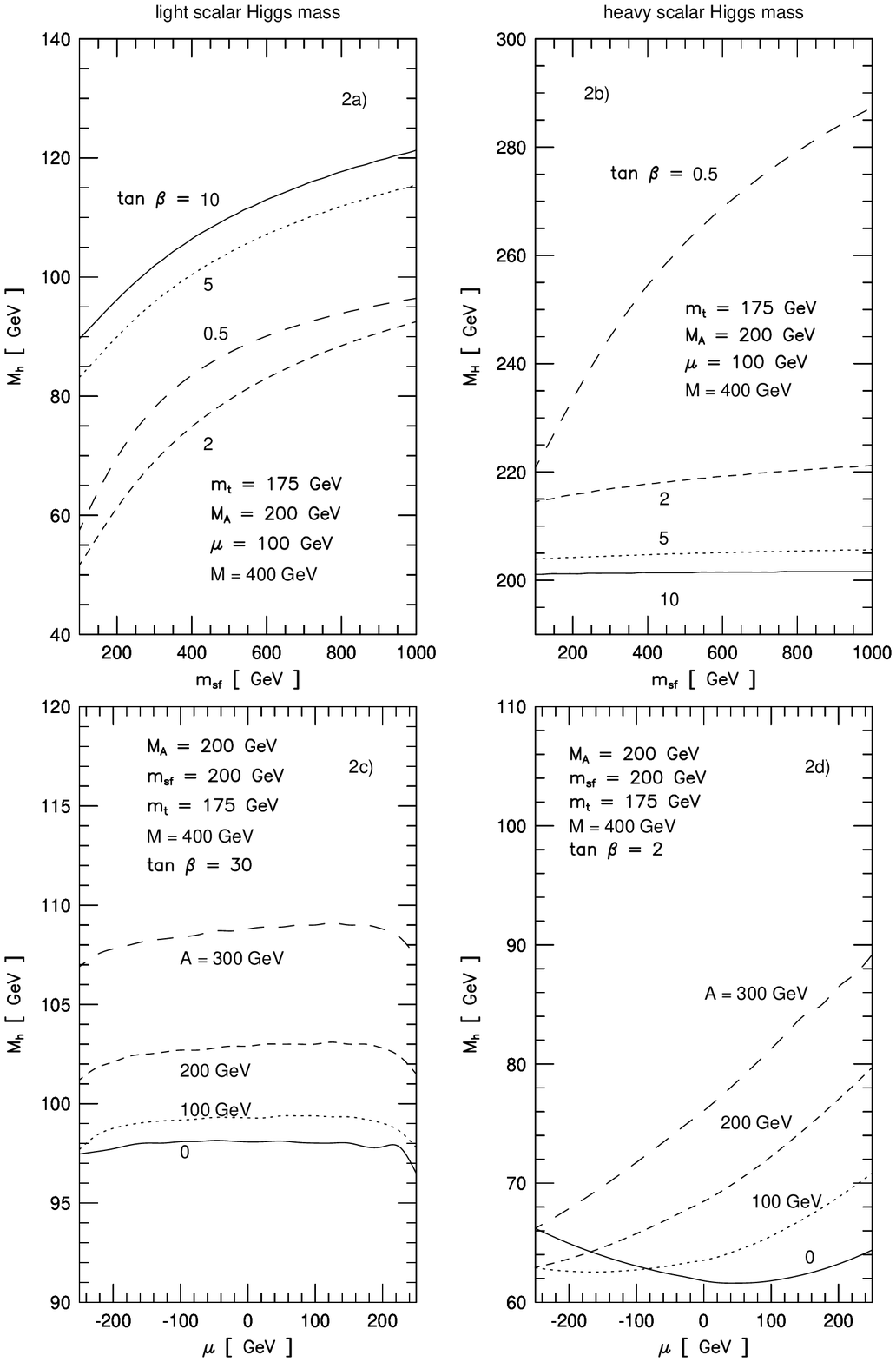,%
height=20cm,width=18.0cm,%
bbllx=75pt,bblly=75pt,bburx=555pt,bbury=751pt}
\caption{}
\end{figure}
\newpage
\begin{figure}
\epsfig{figure=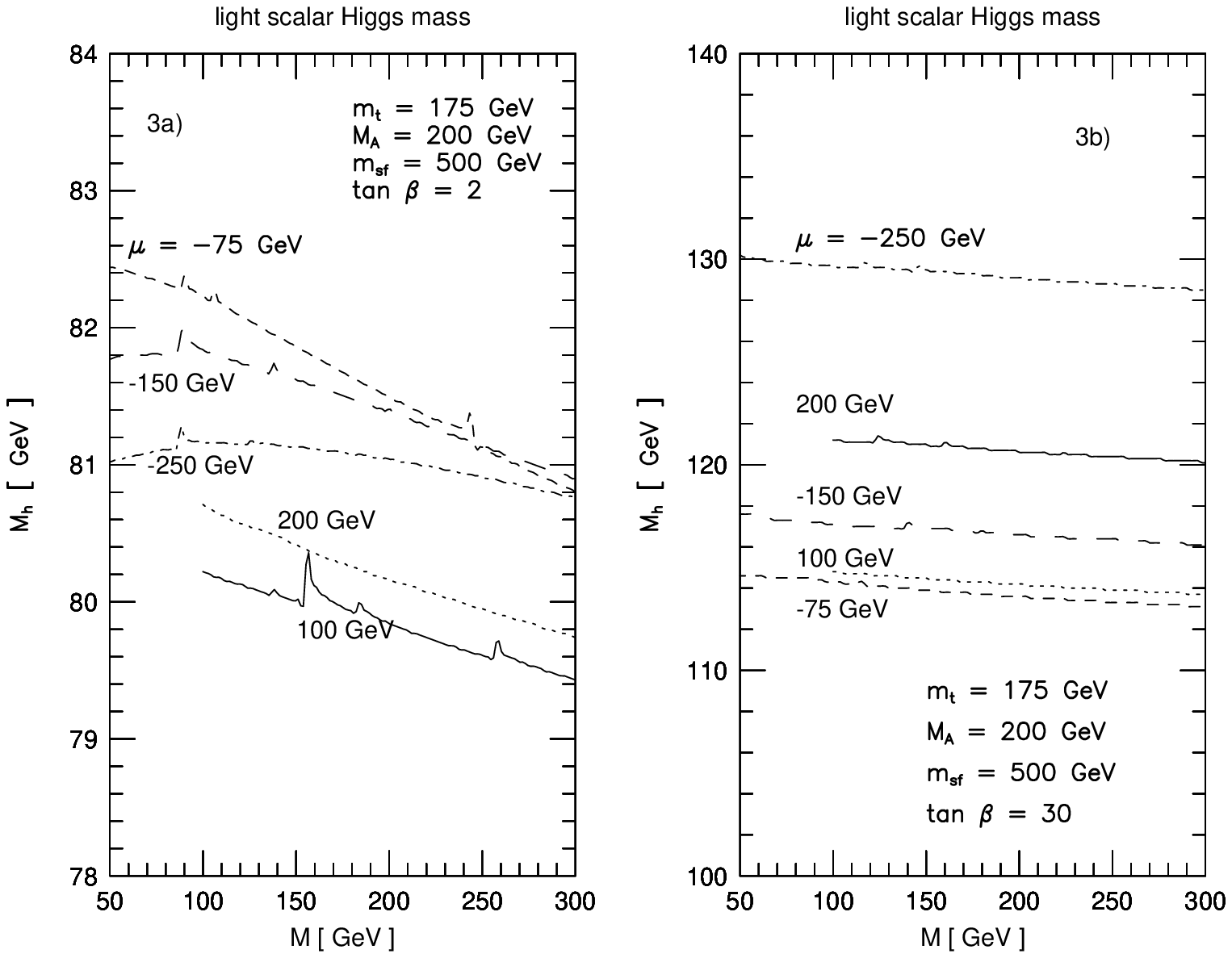,%
height=11cm,width=18.0cm,%
bbllx=75pt,bblly=375pt,bburx=555pt,bbury=751pt}
\caption{}
\end{figure}
\end{document}